\documentclass[aps,onecolumn]{revtex4-1}
\usepackage{hyperref}
\usepackage{graphicx}
\usepackage{float}
\usepackage{amsfonts, amssymb, amsmath, mathrsfs, cases, mathtools, bbm}

\DeclareMathOperator*{\argmin}{arg\,min}
\newcommand{\bx}{\mathbf x}
\newcommand{\bt}{\mathbf t}
\newcommand{\bM}{\mathbf m}
\newcommand{\dd}{\partial}
\newcommand{\sm}{\setminus}
\newcommand{\pr}{\mathbbm P}

\begin{document}

\title{Stochastic optimization by message passing}

\author{F.~Altarelli}
\affiliation{Physics Department and Center for Computational Sciences, Politecnico di Torino, Corso Duca degli Abruzzi 24, 10129 Torino, Italy}
\affiliation{Collegio Carlo Alberto, Via Real Collegio 30, 10024 Moncalieri, Italy}
\author{A.~Braunstein}
\affiliation{Physics Department and Center for Computational Sciences, Politecnico di Torino, Corso Duca degli Abruzzi 24, 10129 Torino, Italy}
\affiliation{Human Genetics Foundation, Torino, via Nizza 52, 10126 Torino, Italy}
\affiliation{Collegio Carlo Alberto, Via Real Collegio 30, 10024 Moncalieri, Italy}
\author{A.~Ramezanpour}
\affiliation{Physics Department and Center for Computational Sciences, Politecnico di Torino, Corso Duca degli Abruzzi 24, 10129 Torino, Italy}
\author{R.~Zecchina}
\affiliation{Physics Department and Center for Computational Sciences, Politecnico di Torino, Corso Duca degli Abruzzi 24, 10129 Torino, Italy}
\affiliation{Human Genetics Foundation, Torino, via Nizza 52, 10126 Torino, Italy}
\affiliation{Collegio Carlo Alberto, Via Real Collegio 30, 10024 Moncalieri, Italy}

\begin{abstract}
Most optimization problems in applied sciences realistically involve uncertainty in the 
parameters defining the cost function, of which only statistical information is known beforehand.
Here we  provide an in-depth discussion of how message passing algorithms for stochastic optimization  based on
the cavity method of statistical physics can be constructed.
We focus on two basic problems, namely the independent  set problem and the matching problem, for which we display the the general method and caveats for the case
of the so called  two-stage problem with  independently distributed stochastic parameters. 
We compare  the results with some greedy algorithms and briefly discuss the extension to more 
complicated stochastic multi-stage problems.

\end{abstract}
\pacs{ } \maketitle

\section{Introduction}\label{sec:intro}
Most real-world optimization problems involve uncertainty: the precise value of
some of the parameters defining the cost function is often unknown, either
because they are measured with insufficient accuracy, or because they are
stochastic in nature and revealed only \emph{after} part of the decisions have been
taken. The purpose of the optimization process  is thus to find solutions which
are optimal in some probabilistic sense, a fact which introduces fundamental
conceptual and computational challenges \cite{papadimitriou2003computational}.

Examples of stochastic optimization problems can be found in all areas of
applied and natural sciences, ranging from resource allocation and robust
design problems in economics and engineering, to problems in chemistry, physics
and biology.  For instance, resilience of biological systems with respect to
unpredictable environmental conditions can be seen as a stochastic optimization
feature selected by evolution, both at the molecular and systems levels. 

Optimization under uncertainty, or Stochastic Optimization, is an ample and
well established field of research which tries to generalize the optimization
methods used in Operations Research and computer science to a probabilistic
setting. The typical framework considered is Two-Stage Stochastic Optimization
(TSSO), in which some of the variables have to be assigned before the
stochastic parameters are specified, and the remaining variables are assigned
after.

TSSO poses some very tough computational challenges: typically, the size of the
uncertainty space is huge (it increases exponentially with the size of the
instance), and the underlying problem  can be computationally hard due to discrete nature of the variables involved (typically decision variables). 
In order to cope with these difficulties, traditional approaches rely on sampling (e.g. considering
``scenarios'') and on the relaxation of the integer constraints.  For example,
Stochastic Programming is the extension of Linear Programming and sampling
techniques to uncertain scenarios \cite{prekopa1995stochastic, Birge, Shapiro}.
In fact, the presence of uncertainty has a deep impact on the computational
complexity of a problem: stochastic optimization problems often belong to a
superset of the NP complexity class called PSPACE
\cite{papadimitriou2003computational}, and many problems which are easy to
solve when their inputs are known exactly, become intractable as soon as some
form of uncertainty is introduced \cite{Kong}.

Statistical mechanics has played  an important role in the past in the design
of large scale optimization algorithms.  Partly this was made possible by
extending ideas from the statistical physics of disordered systems to
applications in computer science. Examples range from Monte Carlo sampling and
simulated annealing \cite{motwani, kirkpatrick}, to the  more recent advances
in message-passing algorithms
\cite{Kabashima,braunstein2003polynomial,neirotti2005improved,
montanari2005compute, chertkov2006loop,
zdeborova2007phase,frey2007clustering,braunstein2008estimating,
bayati2008statistical, ricci2009cavity}.
Monte Carlo can in principle be used to solve TSSO problems as well. What is
needed is to approximate the expectation of a cost function by sampling from
the space of stochastic parameters. Then a second Monte Carlo scheme (e.g.
simulated annealing) is employed to find an optimal solution to the estimated cost
function. This is normally a very heavy computation even for moderately large problem
sizes, especially when stochastic parameters are not concentrated and the
estimated cost function has a complex behavior.

In Ref. \cite{prl} we
introduced an approach to TSSO which resembles Survey Propagation (SP)
\cite{mezard2002random, braunstein2004survey, braunstein2005survey}, combining
Belief Propagation (BP) and its  $\beta=\infty$  version, also known as Max Sum
(MS, see e.g. \cite{Mezard-Montanari} for an extensive review on these two
methods). The approach is partly analytic and it allows to build an algorithm to
optimize the expectation of a stochastic cost function by estimating the
statistics of its minima, without resorting to explicit (and costly) sampling.
The method was applied to a stochastic version of the matching problem with 
independently distributed stochastic parameters and 
it resulted in a distributed message passing algorithm dealing with this TSSO 
problem. The algorithm was shown to perform very well in a stochastic 
bipartite matching problem by extensive numerical simulation.

In this work we discuss the details and the generalization of the method and of the resulting algorithm.
Beside the application to the  stochastic bipartite matching problem introduced in \cite{prl},  we generalize the technique to deal with
a relevant TSSO version of the Maximum Independent Set problem,  which consists in finding the maximum independent set
in a graph, when the node's contribution to the total weight is uncertain
but its distribution is known.  This is a problem that could arise in a
communication network with some interference constraints \cite{Shah2008}.
Consider a network of devices communicating with a central server with the
following constraint: two neighboring devices in this interfrence network can
not transmit information at the same time because the data may be lost.
Therefore, if a set of devices transmit at the same time, they need to be an
independent set of the network to have a successful transmission. In addition,
suppose that the devices have different transmission rates and the server wants
to choose an independent set of the network with maximum transmission rate.
This is a maximum weight independent set problem  \cite{Sanghavi2009}.
If there is a sort of uncertainty in the problem, for example in the
transmission rates, we have to deal with a stochastic optimization problem.   

The two-stage problem can easily be generalized to a multi-stage problem where
in each stage some of the stochastic parameters are revealed and one has to
assign a subset of the variables. This is a more difficult and less studied
problem in the field. 
The method presented here can in principle be extended to study stochastic
multi-stage problems. This generality comes at a cost: every successive stage involves 
dealing with increasingly complex distributions.
In this paper we also consider an heuristics obtained by reducing a multi-stage
problem to a sequence of two-stage problems that seemingly gives a very good
approximation in this particular case.

The paper has the following structure: we define the problem in section
\ref{sec:problem_def}; in section \ref{sec:general_approach} we give the
general cavity approach to solve a two-stage stochastic problem; in section
\ref{sec:stoch_indep_set} we apply the method to the stochastic independent set
problem; finally, in section \ref{sec:matching} we apply it to the stochastic
matching problem.

\section{Problem definition}
\label{sec:problem_def}

In general terms, the problem we study is defined by an energy function
$\mathcal E(\bx_1, \bt_2, \bx_2)$ depending on two sets of decision variables
$\bx_1,\bx_2$ and a set of independent stochastic parameters $\bt_2$.  The
objective is to optimize the average outcome of the following process: first
$\bx_1$ is chosen, then $\bt_2$ is extracted, and then $\bx_2$ is (optimally)
chosen.

That is, the first step consists in fixing $\bx_1$ such that the following
average energy is minimized: \begin{equation}\label{two-stage-1} \bx_1^* =
\argmin_{\bx_1} \mathbb E_{\bt_2} \min_{\bx_2} \mathcal E(\bx_1, \bt_2, \bx_2).
\end{equation}

In certain cases, a greedy algorithm may solve the above problem by replacing
the second stage stochastic parameters with their expected values, that is
\begin{equation}
\label{two-stage-2} \bx_1^{\text{greedy}} = \argmin_{\bx_1}\min_{\bx_2}
\mathcal E(\bx_1, \left<\bt_2\right>, \bx_2).
\end{equation} 
Once the stochastic parameters
$\bt_2$ are extracted, it then solves for the second stage variables, given
$\bx_1^{\text{greedy}}, \bt_2$.  This is a simple but very naive algorithm to
solve a two/multi-stage stochastic optimization problem.  On the other hand, a
clear lower bound for the optimal energy is obtained when all stochastic
parameters are known at the beginning. We call this the offline solution
$\bx^{\text{offline}}$ and it is computed by minimizing the whole energy.  In the
following, we shall always compare the results with the above greedy and offline
solutions.

\section{Cavity approach: passing survey of surveys}
\label{sec:general_approach}

Generally speaking, the method we propose consists in computing the chain of operations in
(\ref{two-stage-1}) by performing the minimizations with the help of MS and the expectation
with BP. The scheme we propose consists loosely in following the procedure below:
\begin{enumerate}
\item Call $\mathbbm{1}_{MS}(\mathbf{m},\bt_2)$ the indicator function for the MS equations 
 for the inner minimum in (\ref{two-stage-1}) as a function of MS messages $\mathbf m$, and 
build the following distribution $\mathcal{Q}(\bt_2, \mathbf{m}) \propto P(\bt_2)\mathbbm{1}_{MS}(\mathbf m, \bt_2)$. 
Separately, compute the MS expression 
for the minimum energy $\mathcal E^*(\bx_1,\bt_2;\mathbf{m})$ on a MS fixed point $\mathbf{m}$. This will 
be needed in item \ref{item:3}.
\item Obtain the BP equations for $\mathcal {Q}$, with message vector $\mathbf Q$. These equations
can be considered as SP equations \cite{braunstein2005survey}. 
\item \label{item:3} Treat
the expression for the minimum energy $\mathcal E^* (\bx_1,\bt_2;\mathbf m)$ from the MS of the first step as an
observable and compute an expression $\mathcal E^*(\bx_1;\mathbf Q)$ for its average $\left<\mathcal E^* (\bx_1,\bt_2;\mathbf m)\right>_{\mathbf Q}$ as a
function of the BP messages. Up to here, the variables $\bx_1$ have been considered
constant. 
\item Finally, employ MS again to find the minimum of $\mathcal
E^*(\bx_1;\mathbf Q)$ over both $\mathbf{Q}$ and $\bx_1$, where the messages are constrained by
the BP equations.
\end{enumerate}

We will now explain more in detail how this is done. Consider a system of interacting variables $V=\{i|i=1,\dots,N\}=V_1 \cup V_2$ with
interaction set $E=\{a|a=1,\dots,M\}$.  Suppose that the energy function is 
\begin{equation}
\mathcal E(\bx_1, \bx_2; \bt_2)=\sum_{a} e_a(x_{\partial a};t_a)+\sum_i e_i(x_i;t_i),
\end{equation}
with variables $\{x_i\}$ and local energies $\{e_i,e_a\}$. 
The stochastic parameters $\{t_i,t_a\}$ are independent and obey a product distribution $P(\bt)=\prod_i p_i(t_i)\prod_a p_a(t_a)$. 
Here $\partial a$ denotes the set of variables contributing in the energy
function $e_a$. Similarly, we use $\partial i$ for the set of interactions depending on $x_i$.  
For fixed $\bx_1$ and $\bt_2$, the statistical physics of the $\bx_2$ variables at finite temperature $T_2=1/\beta_2$,
is given by the following partition function:
\begin{equation}
Z_2[\bx_1;\bt_2] = \sum_{\bx_2} e^{-\beta_2 \mathcal E(\bx_1, \bx_2; \bt_2)}.
\end{equation}

In the Bethe approximation we write the corresponding free energy as
\begin{equation}
F_2[\bx_1;\bt_2] = \sum_{a}\Delta F_a+\sum_{i}\Delta F_i- \sum_{(ia), i \in V_2}\Delta F_{ia}.
\end{equation}
The local free energy changes are computed from the cavity marginals $\psi_{i\to a}(x_i)$ and $\psi_{a\to i}(x_i)$
\begin{align}
e^{-\beta_2 \Delta F_a}&=\sum_{x_{\partial a}} e^{-\beta_2 e_a(x_{\partial a};t_a)} \prod_{i \in \partial a}\psi_{i\to a}(x_i),\\ 
e^{-\beta_2 \Delta F_i}&=\sum_{x_i} e^{-\beta_2 e_i(x_i;t_i)} \prod_{a\in \partial i} \psi_{a\to i}(x_i),\\ 
e^{-\beta_2 \Delta F_{ia}}&=\sum_{x_i} \psi_{i\to a}(x_i)\psi_{a\to i}(x_i),
\end{align}
satisfying the BP equations
\begin{align}\label{BP-psi}
\psi_{i\to a}(x_i)&\propto e^{-\beta_2 e_i(x_i;t_i)}\prod_{b\in \partial i \setminus a}\psi_{b\to i}(x_i)\equiv \hat{\psi}_{i \to a},\\ 
\psi_{a\to i}(x_i)&\propto \sum_{x_{\partial a \setminus i}} e^{-\beta_2 e_a(x_{\partial a};t_a)}
\prod_{j\in \partial a \setminus i}\psi_{j\to a}(x_j)\equiv \hat{\psi}_{a \to i}.
\end{align}
Notice that variables in the first set are fixed, so for these variables $\psi_{i\to a}(x)= \delta_{x,x_i}$ and $\Delta F_i=e_i(x_i;t_i)$.
The right-hand side of (\ref{BP-psi}) should be understood as a definition of the functions $\hat \psi_{i \to a}(\{\psi_{b \to i}, \, b \in \dd i \sm a\}, t_i)$ and $\hat \psi_{a \to i}(\{\psi_{j \to a}, \, j \in \dd a \sm i\}, t_a)$. 

The Bethe approximation to the free energy is asymptotically correct as long as
the interaction graph is locally tree-like and
we are in a replica symmetric phase. For the sake of simplicity, we will assume
that this is the case and that the BP equations have a unique fixed point.
Obviously, when this is not true, assuming replica
symmetry breaking and employing the correspondent RSB equations would give 
a more accurate treatment of the system.

To get the minimum energy configuration we need to take the limit $\beta_2 \to \infty$. 
Let us assume that in this limit the BP messages scale as 
$e^{\beta_2 m_{i\to a}(x_i)}=\psi_{i\to a}(x_i)$ and $e^{\beta_2 m_{a\to i}(x_i)}=\psi_{a\to i}(x_i)$, 
defining new cavity messages $m_{i\to a}$ and $m_{a\to i}$. Starting from the BP equations one can easily
derive the so called Max Sum equations 
\begin{align}\label{MS-m}
m_{i\to a}(x_i)&= -e_i(x_i;t_i)+\sum_{b\in \partial i \setminus a} m_{b \to i}(x_i) \equiv \hat{m}_{i \to a},\\  
m_{a\to i}(x_i)&=\max_{x_{\partial a \setminus i}}\left\{-e_a(x_{\partial a}; t_a)+\sum_{j \in \partial a \setminus i} m_{j\to a}(x_j)\right\}\equiv \hat{m}_{a \to i}.
\end{align}
Again for the variables in the first set
\begin{align}
m_{i\to a}=\log \delta (x_i; \cdot),  \hskip0.5cm  i \in V_1.
\end{align}
where $\log(0)=-\infty$.
To fix a second stage variable we need the local MS messages $m_i$ computed as in (\ref{MS-m}),
but including all the neighbors of $i$ in the sum. With our definition of MS messages, $x_i=\arg \max m_i(x)$. 

Notice that still the messages depend on the stochastic parameters $\bt_2$.
As before we assume that for each realization of $\bt_2$ the MS equations have only one fixed point. The statistics
of the MS messages among different realizations is given by the joint probability distributions $Q_{i \to a}(m_{i \to q};t_i)$ and $Q_{a \to i}(m_{a \to i};t_a)$  satisfying the following equations  
\begin{align}\label{QMS}
Q_{i \to a}(m_{i \to a};t_i)&\propto p_i(t_i)\sum_{\{t_b,m_{b \to i}| b \in \partial i \setminus a\}} 
\prod_{b\in \partial i \setminus a} Q_{b \to i}(m_{b \to i};t_b) \delta(m_{i \to a}-\hat{m}_{i \to a}),\\  
Q_{a \to i}(m_{a \to i};t_a)&\propto p_a(t_a)\sum_{\{t_j,m_{j \to a}| j \in \partial a \setminus i\}} 
\prod_{j\in \partial a \setminus i} Q_{j \to a}(m_{j \to a};t_j) \delta(m_{a \to i}-\hat{m}_{a \to i}). 
\end{align}
Then, the marginals over the MS messages are simply obtained by summing over the stochastic variables 
\begin{align}\label{PMS}
P_{i \to a}(m_{i \to a})=\sum_{t_i}Q_{i \to a}(m_{i \to a};t_i)\equiv \hat{P}_{i \to a}, \\ 
P_{a \to i}(m_{a \to i})=\sum_{t_a}Q_{a \to i}(m_{a \to i};t_a)\equiv \hat{P}_{a \to i}.
\end{align} 
Clearly for fixed first-set variables we have
\begin{equation}
P_{i \to a}(m_{i \to a})=\delta(m_{i \to a}-\log \delta (x_i; \cdot)),  \hskip0.5cm  i \in V_1. 
\end{equation}
We will refer to the above equations as the BP-MS equations. 
The average energy can be computed using the Bethe free energy, that is
\begin{align}
\mathcal E_1(\bx_1)= \sum_{a} \langle \Delta e_a \rangle
+\sum_i \langle \Delta e_i \rangle-\sum_{(ia),i\in V_2} \langle \Delta e_{ia} \rangle, 
\end{align}
where the average of $\Delta e_a=\lim_{\beta_2 \to \infty} \Delta F_a$, $\Delta e_i=\lim_{\beta_2 \to \infty} \Delta F_i$ and $\Delta e_{ia}=\lim_{\beta_2 \to \infty} \Delta F_{ia}$ are taken over the stochastic variables $\bt_2$.

We should mention here that when there exist many Max Sum fixed points the above average energy is computed 
with a uniform measure over the fixed points. Suppose we have $\mathcal{N}_{\bt_2}$ Max Sum fixed points for given $\bx_1, \bt_2$. 
For any fixed point $\mathbf m_{\bt_2}$, consider the Bethe minimum energy $\mathcal E_1(\mathbf m_{\bt_2})$. 
The average energy computed by the surveys $P_{i\to a}(m_{i\to a})$ and $P_{a\to i}(m_{a\to i})$ is indeed 
\begin{align}
\mathcal E'_1(\bx_1)= \sum_{\bt_2} P(\bt_2) \left(\frac{1}{\mathcal{N}_{\bt_2}}\sum_{\mathbf m_{\bt_2}} \mathcal E_1(\mathbf m_{\bt_2})\right). 
\end{align}
It should be noted that, in the case of multiple fixed points, this expression is different the one that would
have been obtained with the loose procedure described at the beginning of this section. Indeed that would have resulted in
\begin{align}
\mathcal E_1''(\bx_1)= \frac{\sum_{\bt_2} P(\bt_2) \left(\sum_{\mathbf m_{\bt_2}} \mathcal E_1(\mathbf m_{\bt_2})\right)}{\sum_{\bt_2} P(\bt_2) \mathcal{N}_{\bt_2}}, 
\end{align}
which is the expression corresponding to item \ref{item:3} of the description.
In particular, $\mathcal E_1''$ may not coincide with $\mathcal E_1$ even in the case in which all the fixed points have the same energy. 
Besides being less informative for our purposes, the computation of $\mathcal E_1''$ is more involved than the one of $\mathcal E'_1$, needing the propagation of joint messages 
$P_{i\to a}(m_{i\to a},m_{a\to i})$ and $P_{a\to i}(m_{a\to i},m_{i\to a})$. In the case of a single fixed point, 
such messages simplify as they depend only on the argument of the forward direction.

Now we are ready to deal with the first stage variables. The partition function for this subsystem at finite temperature $T_1=1/\beta_1$ reads
\begin{equation}
Z_1 = \sum_{\bx_1} e^{-\beta_1 \mathcal E_1(\bx_1)}.
\end{equation}
We recall that the MS messages needed to compute $\mathcal E_1(\bx_1)$ depend implicitly on $\bx_1$. In order to make
the average energy a local function, we introduce the $P_{i \to a}(m_{i \to a})$ and $P_{a \to i}(m_{a \to i})$ 
as new variables in the partition function
\begin{equation}
Z_1 = \sum_{\bx_1,\{P_{i \to a},P_{a \to i}\}} e^{-\beta_1 \mathcal E_1(\bx_1)}
\prod_{i \in V_2}\prod_{a \in \partial i} \delta(P_{i \to a}-\hat{P}_{i \to a}) 
\prod_{a}\prod_{i \in \partial a} \delta(P_{a \to i}-\hat{P}_{a \to i}).
\end{equation}
As before, the marginals of $P_{i \to a}$ and $P_{a \to i}$ can be computed by the Bethe approximation. 
Let us first write the cavity messages related to the second set variables:
\begin{align}\label{BP-Psi-2}
\Psi_{i\to a}(P_{i \to a})&\propto \sum_{\{P_{b \to i}| b\in \partial i \setminus a\}} e^{-\beta_1 
\langle \Delta e_{i\to a}\rangle } 
\prod_{b\in \partial i \setminus a}  \Psi_{b\to i}(P_{b \to i})\delta(P_{i \to a}-\hat{P}_{i \to a}),  \\ 
\Psi_{a\to i}(P_{a \to i})&\propto \sum_{\{P_{j \to a}| j\in \partial a \setminus i\}} e^{-\beta_1 \langle \Delta e_{a\to i}\rangle} 
\prod_{j\in \partial a \setminus i}  \Psi_{j\to a}(P_{j \to a})\delta(P_{a \to i}-\hat{P}_{a \to i}),
\end{align}
where $\langle \Delta e_{i\to a}\rangle$ and $\langle \Delta e_{a\to i}\rangle$ are the average of cavity energy shifts,
\begin{align}\label{Deia}
\Delta e_{i \to a}&=\lim_{\beta_2 \to \infty} \Delta F_{i\to a}, \hskip1cm \Delta F_{i\to a}\equiv \Delta F_{i}-\Delta F_{ia},\\ 
\Delta e_{a \to i}&=\lim_{\beta_2 \to \infty} \Delta F_{a\to i}, \hskip1cm \Delta F_{a\to i}\equiv \Delta F_{a}-\Delta F_{ia}.
\end{align}
Note that the energy term $\Delta e_{i \to a}$ (resp. $\Delta e_{a \to i}$) in (\ref{BP-Psi-2}) does not depend on the backward message $P_{a\to i}$ (resp. $P_{i\to a}$). This is exactly the reason for the regrouping of the energy terms in (\ref{Deia}), and it is crucial in order to avoid correlations between messages traveling in opposite directions.
The cavity messages for the variables in the first set are a bit different from the above equations,
due to the asymmetric form of the average energy,
\begin{align}\label{BP-Psi-1}
\Psi_{i\to a}(x_i)&\propto  e^{-\beta_1 e_i(x_i;t_i)}
\prod_{b\in \partial i \setminus a}  \Psi_{b\to i}(x_i), \\
\Psi_{a\to i}(x_i)&\propto \sum_{\{x_j|j\in \partial a \setminus i, V_1\},\{P_{j \to a}|j\in \partial a \setminus i, V_2\}} e^{-\beta_1 \langle \Delta e_a\rangle} 
\prod_{j\in \partial a \setminus i,V_1}  \Psi_{j\to a}(x_j)\prod_{j\in \partial a \setminus i}  \Psi_{j\to a}(P_{j \to a}).
\end{align}

These finite temperature equations (BP-BP-MS) could already be used to extract useful information about the phase space of the problem. However, in order to find the configuration $\bx_1^*$ minimizing the average energy, we have to take the zero temperature limit $\beta_1 \to \infty$.
Again we work with the following scaling: $\Psi_{i\to a}=e^{\beta_1 M_{i\to a}}$ and $\Psi_{a\to i}=e^{\beta_1 M_{a\to i}}$.
In this way we obtain the Max Sum equations (MS-BP-MS) in the top layer
\begin{align}\label{MS-Mia}
M_{i\to a}(x_i) & = -e_i(x_i;t_i)+
\sum_{b\in \partial i \setminus a}  M_{b\to i}(x_i) \hskip0.5cm i \in V_1, \\ 
M_{i\to a}(P_{i \to a}) &= \max_{\{P_{b \to i}| b\in \partial i \setminus a\}:P_{a\to i}=\hat{P}_{a\to i} }
\left\{ - \langle \Delta e_{i\to a}\rangle+ 
\sum_{b\in \partial i \setminus a}  M_{b\to i}(P_{b \to i}) \right\}\hskip0.5cm i \in V_2,  
\end{align}
and
\begin{align}\label{MS-Mai}
M_{a\to i}(x_i) &= \max_{\substack{\{x_j|j\in \partial a \setminus i, V_1\}, \\ \{P_{j \to a}| j\in \partial a \setminus i\}}} \left\{- \langle \Delta e_a \rangle+ 
\sum_{j\in \partial a \setminus i, V_1}  M_{j\to a}(x_j)+\sum_{j\in \partial a \setminus i, V_2}  M_{j\to a}(P_{j \to a})\right\}\hskip0.5cm i \in V_1,\\ 
M_{a\to i}(P_{a \to i}) &= \max_{\substack{\{x_j|j\in \partial a \setminus i, V_1\},\\ \{P_{j \to a}|j\in \partial a \setminus i, V_2\} : P_{i\to a}=\hat{P}_{i\to a}}}
\left\{ - \langle \Delta e_{a\to i}\rangle+ 
\sum_{j\in \partial a \setminus i, V_1}  M_{j\to a}(x_j)+\sum_{j\in \partial a \setminus i, V_2}  M_{j\to a}(P_{j \to a})\right\}\hskip0.5cm i \in V_2. 
\end{align}

The above messages should be normalized by subtracting the maximum value of the unnormalized message in each case.
Starting from random initial messages $\{M_{i\to a},M_{a\to i}\}$ we update them according to the above equations.
At the fixed point the local messages $M_i$ determine the solution to the first stage variables. Introducing a small reinforcement to the equations
would help the algorithm converge more easily to a polarized solution \cite{braunstein2006learning}. 
To this end we modify a bit the equations for the first set variables as
\begin{align}\label{rMS-Mij}
M_{i\to a}(x_i)= - e_i(x_i;t_i) +\sum_{b\in \partial i \setminus a} M_{b\to i}(x_i)+ \rho M_i(x_i), \\ 
M_{i}(x_i )= - e_i(x_i;t_i) +\sum_{b\in \partial i} M_{b\to i}(x_i) + \rho M_i(x_i),
\end{align}
where $\rho \ge 0$ is the reinforcement parameter.

In the next sections we will make the above points more clear by studying two problems: a stochastic independent set problem and a stochastic matching problem.

\section{The two-stage stochastic independent set problem}
\label{sec:stoch_indep_set}

We consider a weighted graph $G=(V,E)$ with node set $V=V_1\cup V_2$ of size $N$, edge set $E$ and weights $w_i$ on the nodes $i=1,\ldots,N$.
A configuration of nodes $\bx \in \{0,1\}^N$ defines an independent set if $x_i x_j=0$ for any edge in $E$. 
The weight of an independent set is the weight of nodes belonging to the set, i.e. $W=\sum_i x_iw_i$, and a maximum independent
set has the maximum weight among all the independent sets. A stochastic version of this problem is obtained by introducing independent 
stochastic parameters $t_i\in \{0,1\}$ representing the nodes that contribute to the total weight.
Given the probability distributions $\{p_i(t_i)|i \in V_2\}$ we are to find an independent set with maximum
weight $W=\sum_i t_ix_iw_i$, after realizing $\bt_2$. In terms of the previous section notations: 
$e_i(x_i;t_i)=-t_i x_i w_i$ and $e_a \equiv e_{ij}(x_i,x_j)=\delta_{x_ix_j,1} \times \infty $. That is we
have deterministic hard interactions and no stochastic parameters $t_a$.

In the following we shall work with Erdos-Renyi (ER) random graphs. The subsets $V_1$ and $V_2$ are chosen randomly:
a node can belong to the first or the second subset with equal probability. For the sake of simplicity we shall
assume that all the node weights are the same, say $w_i=1$ for any $i$.
The probability distributions $p_i(t_i)=p_i \delta_{t_i,1}+(1-p_i) \delta_{t_i,0}$ define the amount of uncertainty in the problem. 
When all $p_i=0$ or $p_i=1$, there is no uncertainty and we recover the deterministic problem. 
In the other extreme we have all $p_i=1/2$ that is the most uncertain case.

\subsection{Message passing solution}
For a fixed $\bx_1$ which is an independent set of the subgraph induced by $V_1$, we write
\begin{align}
Z_2[\bx_1;\bt_2]=\sum_{\bx_2} \prod_{(ij)\in E} \delta_{x_ix_j,0} e^{\beta_2 \sum_i t_i x_i }.
\end{align}
The BP equations for this problem are
\begin{align}
\psi_{i\to j}(x_i=0) &\propto 1,\\ 
\psi_{i\to j}(x_i=1) &\propto e^{\beta_2 t_i}\prod_{k\in \partial i \setminus j}\psi_{k\to i}(x_k=0),
\end{align}
which can be understood as messages from a variable to a constraint. 
The equations converge on an ER random graph for any  $\beta_2<\beta_2^*$, which depends on the average connectivity. 
For smaller temperatures the replica symmetry assumption is not anymore correct. 
Then we obtain Max Sum equations, which for binary variables simplify slightly as messages can be parametrized 
with a single real number $m(1)-m(0)$:
\begin{align}\label{IS-MSij-x}
m_{i\to j}&=m_i=(2x_i-1)\times \infty \hskip0.5cm i\in V_1,\\ 
m_{i\to j}&=t_i-\sum_{k\in \partial i\setminus j}\max(0,m_{k\to i})\equiv \hat{m}_{i\to j} \hskip0.5cm i\in V_2.
\end{align}
These equations converge on an ER random graph as long as the average degree is smaller than $\exp(1)$. 
In the following we will always use these equations to find, for example, the greedy and offline solutions.
To improve the convergence the algorithm for large degrees one can introduce very small noises in the weights $w_i$ and use the reinforced equations.

The distributions of the MS messages over stochastic parameters $\bt_2$ are given by
\begin{equation}
P_{i\to j}(m_{i\to j})\propto
\sum_{t_i} p_i(t_i) \sum_{\{m_{k \to i}| k \in \partial i \setminus j\}} 
\prod_{k\in \partial i\setminus j} P_{k\to i}(m_{k\to i})\delta(m_{i\to j}-\hat{m}_{i\to j})\equiv \hat{P}_{i\to j}.
\end{equation}
The equations for $P_{i\to j}(+1)$ are simply written as   
\begin{align}
P_{i\to j}(+1)=p_i\prod_{k \in \partial i \setminus j}\left(1-P_{k\to i}(+1)\right),
\end{align}
with no need of the other probabilities. The normalization condition gives the probability of having zero and negative messages.  
The above survey can be used to compute the average energy $\mathcal{E}_1(\bx_1)$ for a given configuration of the first stage variables:
\begin{equation}
\mathcal{E}_1(\bx_1)= \sum_{i\in V_1} t_i x_i
+\sum_{i\in V_2} \langle \Delta e_i \rangle-\sum_{(ij) \in E, (i,j) \in V_2} \langle \Delta e_{ij} \rangle.
\end{equation}
The average energy shifts are 
\begin{align}
\langle \Delta e_i \rangle &=\sum_{m_{i}> 0}  m_{i} P_{i}(m_{i})=P_{i}(m_{i}=+1), \\ 
\langle \Delta e_{ij}\rangle &=\langle \Delta e_i \rangle-\langle \Delta e_{i\to j}\rangle,
\end{align}
with
\begin{align}
\langle \Delta e_{i\to j}\rangle=\sum_{m_{i\to j}> 0}  m_{i\to j} P_{i\to j}(m_{i\to j})=P_{i\to j}(m_{i\to j}=+1).
\end{align}

\begin{figure}
\includegraphics[width=10cm]{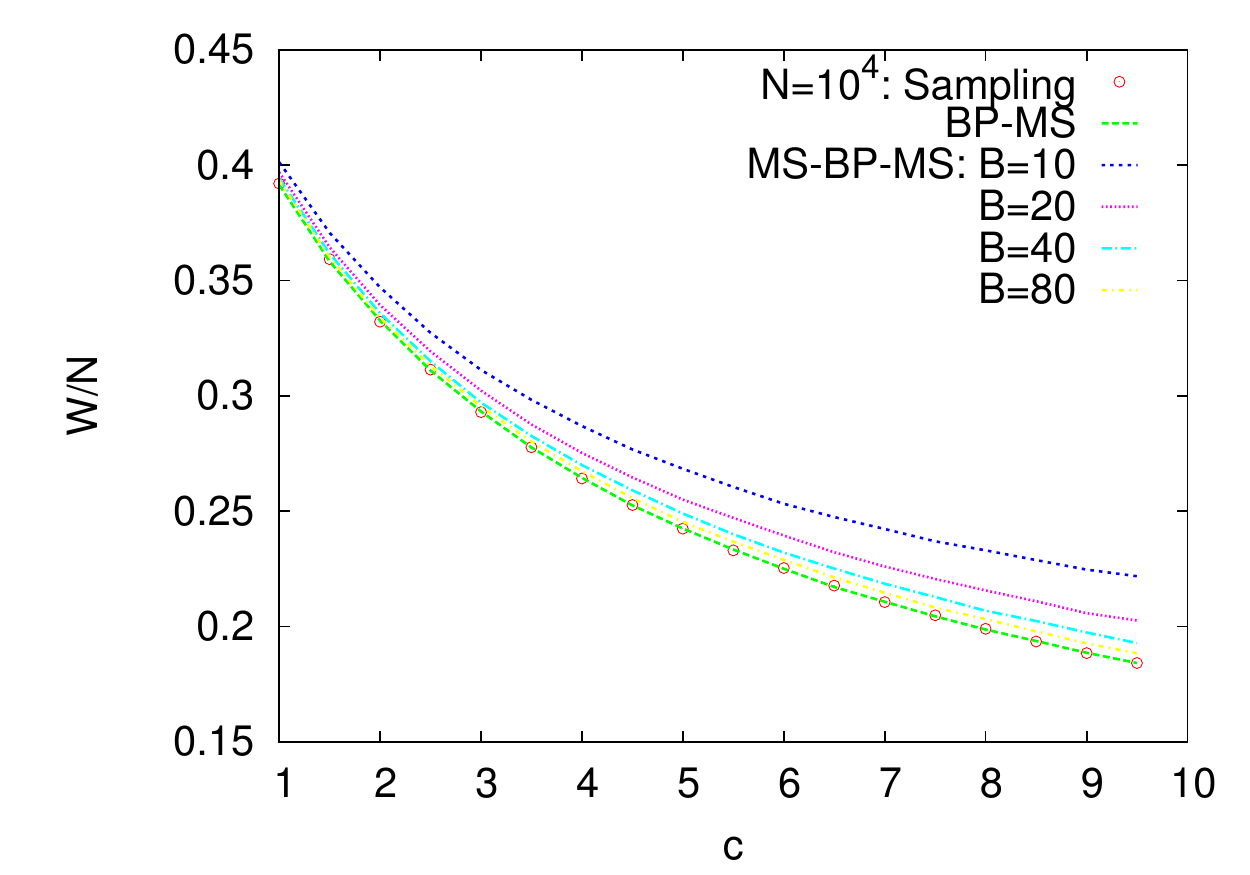}
\caption{The 2-stage problem: the average weight of independent sets, given $\bx_1$, obtained from the surveys and by sampling. The variables in the first set are chosen randomly with
probability $1/2$. The stochastic parameters are in the most uncertain state, i.e. $p_i(t_i=1)=1/2$.
Also the first set nodes contribute to the total weight with probability $1/2$.
The data are the result of averaging over $100$ instances of random graphs, weights and stochastic parameters. The size of the graph is $N=10^4$,
$c$ is the average degree and $B$ denotes the number of bins. The number of bins increases from top to bottom.}\label{W1-2stage}
\end{figure}

Let us compare the above average energy with the one obtained by sampling the stochastic parameters. More precisely, we generate a large number $\mathcal{S}$ of samples $\bt_2^s$
from distribution $\prod_{i\in V_2} p_i(t_i)$ and find the minimum energy configuration $\bx_2^s$ using the Max Sum algorithm. Thus, the average energy can be written as
\begin{equation}
\mathcal{E}_1^{(s)}(\bx_1)=\frac{1}{\mathcal{S}}\sum_s \mathcal{E}_1(\bx_1,\bx_2^s;\bt_2^s).
\end{equation}
In figure \ref{W1-2stage} we compare the two average energies.

At finite temperatures the top layer BP equations read 
\begin{align}\label{IS_BP}
\Psi_{i\to j}(x_i) &\propto \sum_{\{x_k| k\in \partial i \setminus j, V_1\}: x_ix_k=0} 
e^{\beta_1 t_i x_i}\prod_{k\in \partial i \setminus j, V_1} \Psi_{k\to i}(x_k)\prod_{k'\in \partial i \setminus j, V_2} \Psi_{k'\to i}(x_i) \hskip0.5cm i\in V_1,\\ 
\Psi_{i\to j}(x_j) &\propto \sum_{\substack{\{x_k| k\in \partial i \setminus j,V_1\},\\ \{P_{k\to i}| k\in \partial i \setminus j,V_2\}}} 
e^{\beta_1 \langle \Delta e_i \rangle }\prod_{k\in \partial i \setminus j,V_1} \Psi_{k\to i}(x_k)\prod_{k\in \partial i \setminus j,V_2} \Psi_{k\to i}(P_{k\to i}) \hskip0.5cm i\in V_2, j\in V_1, \\ 
\Psi_{i\to j}(P_{i\to j}) &\propto \sum_{\substack{\{x_k| k\in \partial i \setminus j,V_1\},\\ \{P_{k\to i}| k\in \partial i \setminus j\}}} 
e^{\beta_1 \langle \Delta e_{i\to j}\rangle }\prod_{k\in \partial i \setminus j,V_1} \Psi_{k\to i}(x_k)\prod_{k\in \partial i \setminus j} \Psi_{k\to i}(P_{k\to i}) \delta(P_{i\to j}-\hat{P}_{i\to j})  \hskip0.5cm i\in V_2, j\in V_2
\end{align}
Notice that for variables in the first set the sums are restricted by the hard constraints. 
Moreover, since the energy shifts depend only on $P_{i\to j}(+1)$, 
we need just to consider this probability in the equations. Therefore, the relevant variable in the messages is $P_{i\to j}(+1)$.

The top layer MS equations are obtained as before by taking the $\beta_1 \to \infty$ limit in the finite temperature equations:
\begin{align}\label{IS_MS}
M_{i\to j}(x_i) &=\max_{\{x_k|k\in \partial i\setminus j, V_1\}: x_ix_k=0}
\left\{t_i x_i+\sum_{k\in \partial i\setminus j,V_1} M_{k\to i}(x_k)+\sum_{k'\in \partial i\setminus j,V_2} M_{k'\to i}(x_i)\right\} \hskip0.5cm i\in V_1, \\ 
M_{i\to j}(x_j) &=\max_{\substack{\{x_k| k\in \partial i \setminus j,V_1\},\\ \{P_{k\to i}|k\in \partial i\setminus j,V_2\}}}
\left\{\langle \Delta e_i \rangle+\sum_{k\in \partial i\setminus j,V_1} M_{k\to i}(x_k)+\sum_{k\in \partial i\setminus j,V_2} M_{k\to i}(P_{k\to i})\right\} \hskip0.5cm i\in V_2, j\in V_1, \\  
M_{i\to j}(P_{i\to j}) &=\max_{\substack{\{x_k| k\in \partial i \setminus j,V_1\},\\ \{P_{k\to i}|k\in \partial i\setminus j,V_2\}:P_{i\to j}=\hat{P}_{i\to j}}}
\left\{\langle \Delta e_{i\to j}\rangle+\sum_{k\in \partial i\setminus j,V_1} M_{k\to i}(x_k)+\sum_{k\in \partial i\setminus j,V_2} M_{k\to i}(P_{k\to i})\right\} \hskip0.5cm i\in V_2, j\in V_2. 
\end{align}

One strategy to solve the above equations is to work with discrete variables taking a small number of values. Let us assume $P_{i\to j}(+1)$ takes $B+1$ discrete values in $[0,1]$, that is $P_{i\to j}(+1)=n/B$ for $n\in \{0,1,\dots,B\}$.
Given this binning, one
could try to solve the equations by summing exhaustively over all the possible configurations of the variables. 
The time complexity of this computation grows exponentially with the degree of nodes as $B^d$. 
We can do better than exhaustive sum by using the distributive nature of the equations. 
When $i$ is in the first set, the input variables $\{x_k|k\in V_1\}$ are decoupled for different neighbors $k$ and we need only to take care of the hard constraints.  When $i$ is in the second set, we have to sum over all the values of the input variables giving rise to the specific output variable $P_{i \to j}(m_{i\to j}=+1)$, which depends only on the product of $(1-P_{k \to i}(m_{k\to i}=+1))$ for different $k\in \partial i\setminus j$ and $p_i(t_i=1)$. This can be done by splitting the whole sum into smaller ones such that at each step we get a convolution of the new messages and the sum over previous messages, that is
\begin{align}
F_{i \to j}^l(\tilde{P})=\max_{P_l}\left\{(F_{i\to j}^{l-1}(\frac{\tilde{P}}{1-P_l}),M_{k_l\to i}(P_l)\right\}, 
\end{align}
where $l$ goes from $1$ to $d_i-1=|\partial i|-1$. In the last step we update the message as
\begin{align}
M_{i\to j}(P_{i\to j})=\langle \Delta e_{i\to j}\rangle+F_{i \to j}^{d_i-1}(\tilde{P}=\frac{P_{i\to j}}{p_i}). 
\end{align}
Now we need $B^2d_i$ operations to update a cavity message.
The time complexity of this algorithm grows linearly with $N$ for finite degree graphs and finite number of bins. In Figure \ref{W1-2stage} we display the average weight of independent sets for a fixed configuration of the first stage variables computed by the above equations with discrete variables.
This is just to be sure that by summing over the discrete surveys in these equations we recover the correct average energy. 

\begin{figure}
\includegraphics[width=10cm]{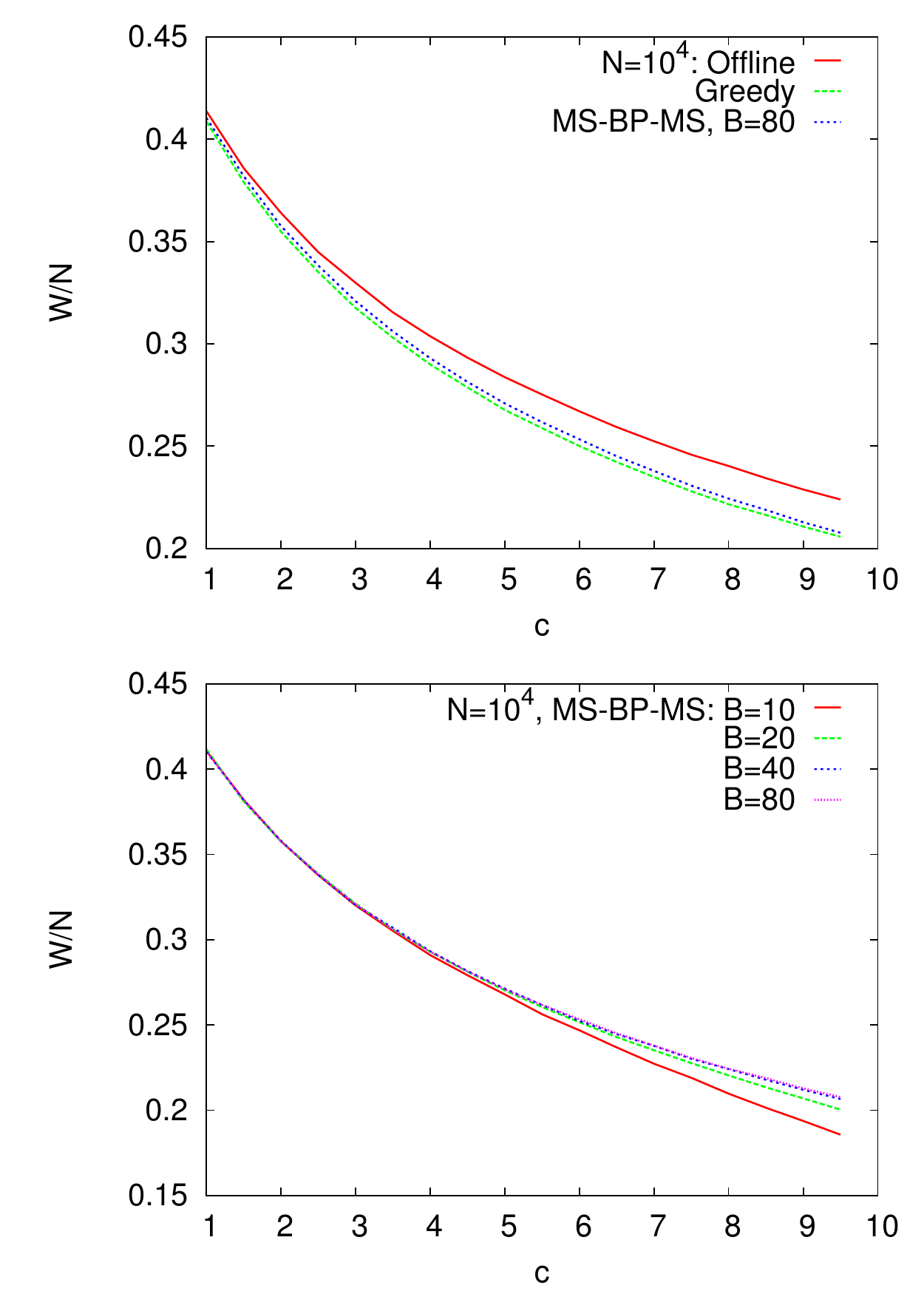}
\caption{The 2-stage problem: comparing the weight of independent sets obtained by the messages passing algorithm with the greedy and offline algorithms. The variables in the first set are chosen randomly with
probability $1/2$. The stochastic parameters are in the most uncertain state, i.e. $p_i(t_i=1)=1/2$.
Also the first set nodes contribute to the total weight with probability $1/2$.
The data are result of averaging over $100$ instances of random graphs, weights and stochastic parameters.
The size of graph is $N=10^4$ and $c$ is the average degree. The number of bins $B$ increases from bottom to top.}\label{W2-2stage}
\end{figure}

However, the above equations are indeed to find the optimal configuration of the first stage variables as described before. Figure \ref{W2-2stage} compares the performance of the algorithm with the greedy and offline solutions. The figure also shows how the maximum weight solutions obtained in this way depend on the number of bins.

\subsection{Monte Carlo approach: sampling + local search}\label{S4}
The two stage stochastic problem can in principle be studied by a Monte Carlo algorithm.
Given a problem instance, we extract $\mathcal{S}$ samples of $\bt_2$ from the probability distribution $\prod_{i\in V_2} p_i(t_i)$.
For a fixed $\bx_1$ and a sample $\bt_2^s$, one finds $\bx_2^s$ that minimize the total energy
\begin{equation}\label{x2-MC}
\bx_2^s = \argmin_{\bx_2} \mathcal E(\bx_1, \bx_2; \bt_2^s).
\end{equation}
Then we find $\bx_1^*$ to minimize the average total energy 
\begin{equation}\label{x1-MC}
\bx_1^* = \argmin_{\bx_1} \frac{1}{\mathcal{S}}\sum_{s=1}^{\mathcal{S}} \mathcal E(\bx_1, \bx_2^s; \bt_2^s).
\end{equation}

\begin{figure}
\includegraphics[width=10cm]{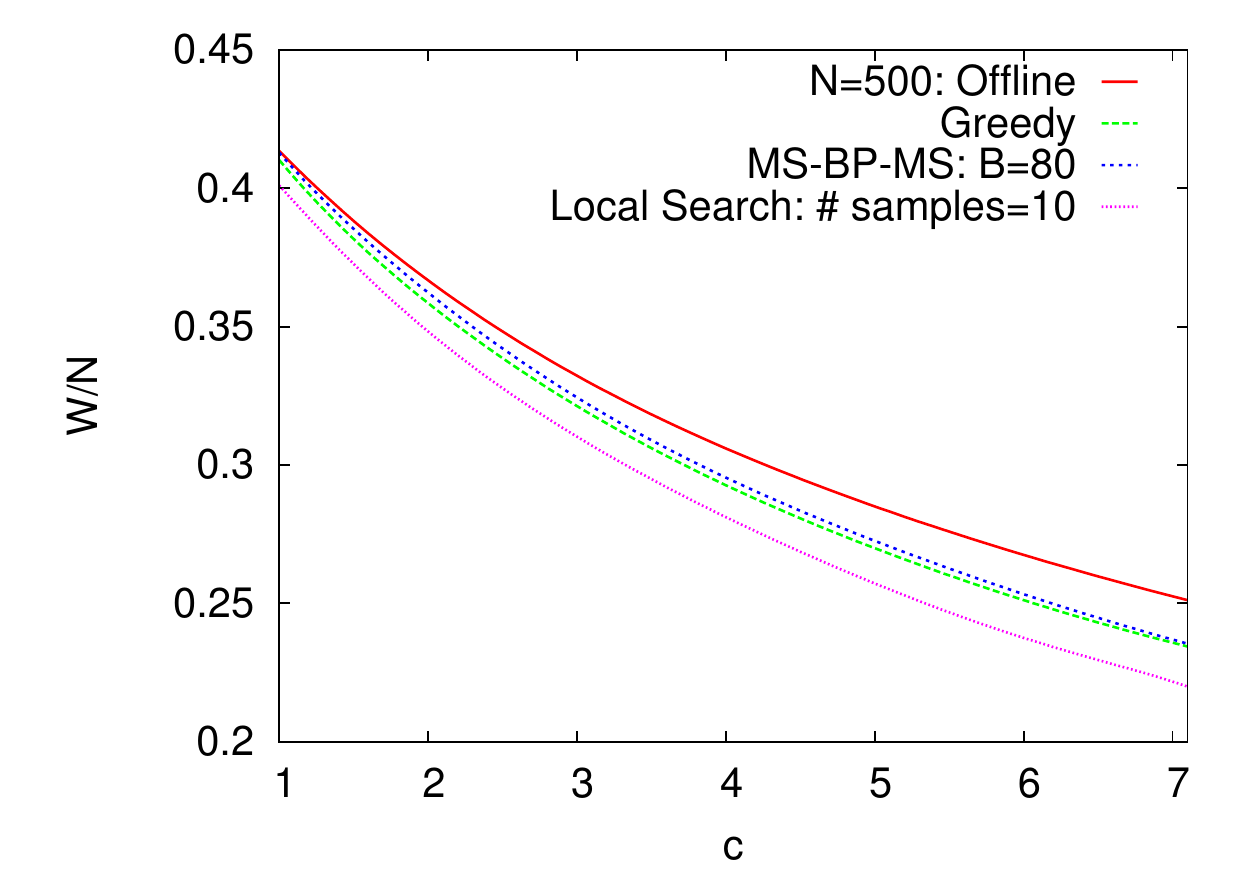}
\caption{The 2-stage problem: comparing the weight of independent sets obtained by the local search algorithm with the other algorithms. The variables in the first set are chosen randomly with
probability $1/2$. The stochastic parameters are in the most uncertain state, i.e. $p_i(t_i=1)=1/2$.
Also the first set nodes contribute to the total weight with probability $1/2$.
The data are result of averaging over $100$ instances of random graphs, weights and stochastic parameters. Size of graph is $N=500$, $c$ is the average degree
and number of samples is $10$.}\label{W2-2stage-MC}
\end{figure}

So far, the only difference with the previous sections is in replacing the average energy with
an average over a finite number of samples. Then we have to choose an algorithm to solve the above two
optimization problems. Here we use a mixture of Max Sum and zero temperature Monte Carlo; Max Sum to find
$\bx_2^s$ and Monte Carlo to find $\bx_1^*$. Given $\{\bt_2^s|s=1,\dots,\mathcal{S}\}$
we start from $\bx_1=0$. Then we select randomly a node $i$ from $V_1$ and flip $x_i$ to $x_i^{new}$. This results to 
a change in the average energy $\Delta \mathcal{E}=\frac{1}{\mathcal{S}}\sum_{s=1}^{\mathcal{S}} \Delta \mathcal{E}^s$. 
Notice that to compute
$\Delta \mathcal{E}^s$ we have to find $\bx_2^s$ and we do this by using the Max Sum algorithm.  
In a zero temperature Monte Carlo we accept the change to $x_i^{new}$ only if $\Delta \mathcal{E}<0$. We repeat the above 
steps until the algorithm finds a local minimum of the average energy function. In an iteration of the algorithm 
all the first set variables are selected in a random sequential way.  
In figure \ref{W2-2stage-MC} we compare the outcome of this algorithm with the greedy and offline solutions.   
The algorithm is computationally expensive, and therefore in order to obtain a good statistic we had to restrict ourself to a small graph. The time complexity of these algorithms increases as $\mathcal{S}N^2$ for finite degree
graphs and when a finite number of iterations are enough to reach a good approximate solution.
Here we assumed that the number of first stage nodes scales with $N$. 
Notice that, instead of zero temperature Monte Carlo we could use a more sophisticated algorithm like 
simulated annealing, but it would be more time consuming.

\subsection{Multi-stage stochastic optimization}\label{S5}
In this section we see how the message passing method can be generalized to study a $K$-stage problem. With obvious notations, the problem at stage $l$
is to find the partial configuration $\bx^*_l$ which minimizes the expected value of the final cost function $\mathcal E(\bx, \bt)$, 
given the previously assigned variables $\bx_1, \dots, \bx_{l-1}$ and the previously set parameters $\bt_1, \dots, \bt_{l}$:
\begin{align}\label{k-stage-1}
  \bx^*_l = \argmin_{\bx_l} \mathbb E_{\bt_{l+1}} \min_{\bx_{l+1}} \cdots 
\mathbb E_{\bt_K} \min_{\bx_K}  \mathcal E(\bx, \bt).
\end{align}
A greedy algorithm solves the problem at stage $l$ by minimizing the total energy function replacing the unknown stochastic variables with their expectations. 
The offline solution is computed by minimizing the whole energy given $\bt_1,\dots,\bt_K$.

Starting form the bottom layer BP equations at temperature $T_K$ we could compute $\Psi^K_{i\to j}$ depending on $\bx_1,\dots,\bx_{K-1}$
and $\bt_1,\dots,\bt_{K}$. These messages contain all we need to know about the variables $\bx_K$. We denote the corresponding Max Sum messages by $M^K_{i\to j}$. The probability of these messages over the stochastic parameters $\bt_K$ is given by $P_{i\to j}^{K}(M^K)$.
We could use these surveys to write the next layer MS equations $M_{i\to j}^{K-1}(P_{i \to j}^{K})$ which give the information necessary  
for fixing variables $\bx_{K-1}$. Similarly we get probabilities over the stochastic parameters $\bt_{K-1}$ in the surveys $P_{i\to j}^{K-1}(M_{i \to j}^{K-1})$
and these give rise to the new set of MS equations $M_{i\to j}^{K-2}(P_{i \to j}^{K-1})$. In summary, at stage $l$ we compute the MS messages as
\begin{align}\label{MS-Mij-l}
M_{i\to j}^l(x_i) &= \max_{\{x_k| k\in \partial i \setminus j,V_{l'\le l}\}:x_ix_k=0} \left\{ t_ix_i w_i+
\sum_{k\in \partial i \setminus j,V_{l'\le l}} M_{k\to i}^l(x_k)+
\sum_{k\in \partial i \setminus j,V_{l'> l}} M_{k\to i}^l(x_i) \right\} \hskip0.5cm i \in V_l, \\ 
M_{i\to j}^l(x_j) &= \max_{\{P_{k \to i}^{l+1}| k\in \partial i \setminus j\}} \left\{ 
-\langle \Delta e_{i}^{l+1} \rangle
+\sum_{k\in \partial i \setminus j} M_{k\to i}^l(P_{k \to i}^{l+1}) \right\} \hskip0.5cm i \in V_{l'>l}, j\in V_l \\ 
M_{i\to j}^l(P_{i \to j}^{l+1}) &= \max_{\{P_{k \to i}^{l+1}| k\in \partial i \setminus j\}:P_{i \to j}^{l+1}=\hat{P}_{i \to j}^{l+1}} \left\{- \langle \Delta e_{i\to j}^{l+1} \rangle
+\sum_{k\in \partial i \setminus j} \Psi_{k\to i}^l(P_{k \to i}^{l+1})\right\} \hskip0.5cm i,j \in V_{l'>l},
\end{align}
where for variables in stage $l$ and before that, the messages $M_{i\to j}^l(P_{i \to j}^{l+1})$ are concentrated on $x_i$.

The messages statistics are given by
\begin{align}
P_{i\to j}^l(M_{i\to j}^l) & \propto \sum_{t_i} p_i(t_i) \sum_{\{M_{k \to i}^l| k \in \partial i \setminus j\}} 
\prod_{k\in \partial i\setminus j} P_{k\to i}^l(M_{k\to i}^l)\delta(M_{i\to j}^l-\hat{M}_{i\to j}^l) \hskip0.5cm i \in V_l, \\ 
P_{i\to j}^l(M_{i\to j}^l) & \propto \sum_{\{M_{k \to i}^l| k \in \partial i \setminus j\}} 
\prod_{k\in \partial i\setminus j} P_{k\to i}^l(M_{k\to i}^l)\delta(M_{i\to j}^l-\hat{M}_{i\to j}^l) \hskip0.5cm i \in V_{l'>l}.
\end{align}
For variables fixed in the previous stages, $P_{i\to j}^l(M_{i\to j}^l)$ is concentrated on $x_i$. 
As before we used $\hat{O}$ to denote the corresponding equation for quantity $O$. 
And finally, the average cavity energies are computed by
\begin{align}
\langle \Delta e_{i}^{l} \rangle &= \sum_{\{M_{k\to i}^l|k\in \partial i\}} \Delta e_{i}^l \prod_{k\in \partial i}P_{k\to i}^l(M_{k\to i}^l),\\
\langle \Delta e_{i\to j}^{l} \rangle &= \sum_{\{M_{k\to i}^l|k\in \partial i \setminus j\}} \Delta e_{i\to j}^l \prod_{k\in \partial i \setminus j}P_{k\to i}^l(M_{k\to i}^l).
\end{align}
where
\begin{align}
\Delta e_{i}^l &= \max_{\{P_{k \to i}^{l+1}| k\in \partial i\}} \left\{- \langle \Delta e_{i}^{l+1} \rangle+
\sum_{k\in \partial i} M_{k\to i}^l(P_{k \to i}^{l+1}) \right\},\\ 
\Delta e_{i\to j}^l &= \max_{\{P_{k \to i}^{l+1}| k\in \partial i \setminus j\}} \left\{- \langle \Delta e_{i\to j}^{l+1} \rangle+
\sum_{k\in \partial i \setminus j} M_{k\to i}^l(P_{k \to i}^{l+1}) \right\}.
\end{align}

\begin{figure}
\includegraphics[width=10cm]{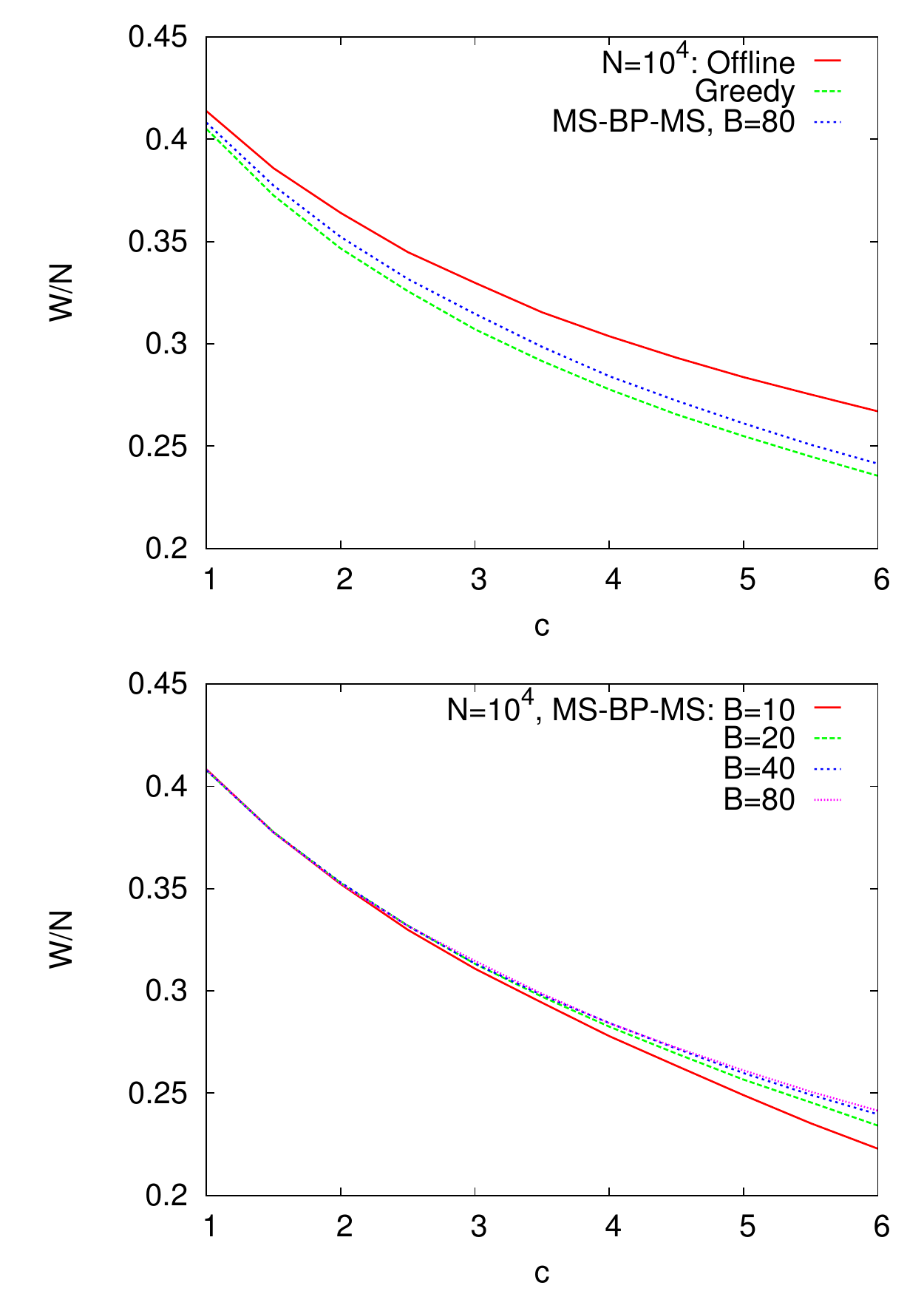}
\caption{The 10-stage problem: comparing the weight of independent sets obtained by the messages passing algorithm with the greedy and offline algorithms.
The variables in each set are chosen randomly with probability $1/10$. The stochastic parameters are in the most uncertain state, i.e. $p_i(t_i=1)=1/2$. Also the first set nodes contribute to the total weight with probability $1/2$.
The data are result of averaging over $100$ instances of random graphs, weights and stochastic parameters.
Size of graph is $N=10^4$ and $c$ is the average degree. Number of bins $B$ increases from down to top.}\label{W2-10stage}
\end{figure}

Notice the nested nature of the messages, which makes an exact treatment of the above equations nearly 
impossible for large $K$.  
However, the efficiency of the algorithm for $K=2$ allows to use it to obtain an approximate solution to the 
$K$-stage problem with $K>2$. A simple heuristics consists in finding
\begin{align}\label{k-stage-2}
\mathbf {\tilde x}_l^* = \argmin_{\bx_l} \mathbb E_{\bt_{l+1} \cdots \bt_K} 
\min_{\bx_{l+1} \cdots \bx_K} \mathcal E(\bx, \bt),
\end{align}
by repeatedly applying the algorithm for a two stage problem. Changing the order of minimization and expectation at each stage, 
we produce lower bounds for the expected value of the energy. As $K$ increases, the approximation effects are accumulated, 
resulting to a suboptimal solution. However, as figure \ref{W2-10stage} shows, we get still 
better results than the greedy algorithm.

\section{The two-stage stochastic matching problem}
\label{sec:matching}

As a second illustration of the method described in Section \ref{sec:general_approach}, let us consider the following problem, which is a variant of the stochastic two-stage bipartite matching problem introduced in \cite{Kong, Katriel, Escoffier}, where it is shown to be NP-complete, and for which the main results have already been published in \cite{prl}. We are given a bipartite graph $G = (L, R; E)$ with $L$ further partitioned in $L_1$ and $L_2$, and for each $l_2 \in L_2$ a real number $p_{l_2} \in ]0,1[$. The objective is to find a maximum-size matching under the following two-stage setup: the vertices in $L_1$ are deterministic, and they must be matched in the first stage of the problem; the vertices in $L_2$ are stochastic, i.e. they may or may not be available for matching, and the available ones must be matched in the second stage. After the first-stage vertices have been matched, the available vertices for the second stage are extracted (independently for each vertex $l_2 \in L_2$ with probability $p_{l_2}$), and then the second-stage optimization is performed. Therefore the optimization in the first stage must be done knowing only partial information (i.e. the probabilities $\mathbf p = \{p_{l_2},\, l_2 \in L_2\}$) about the availability of the second-stage vertices, and once the available vertices in $L_2$ are known in the second stage, the matching of the first-stage vertices cannot be modified.

We introduce two sets of binary variables, $\bx_1 = \{x_{l_1 r} \in \{0,1\}, (l_1 r) \in E : l_1 \in L_1\}$ and $\bx_2 = \{x_{l_2 r} \in \{0,1\}, (l_2 r) \in E : l_2 \in L_2\}$, to represent the possible $M \subset E$, with $x_{lr} = 1$ if and only if $(lr) \in M$, and a set of binary stochastic parameters $\bt = \{t_{l_2} \in \{0,1\}, l_2 \in L_2\}$ with $t_{l_2}=1$ if and only if $l_2$ is available for matching in the second stage, so that $\pr[t_{l_2} = 1] = p_{l_2}$ (notice a slight change in the notation compared to the previous section, which makes it more suitable for this specific problem). We define an energy function $\mathcal E(\bx_1, \bt, \bx_2)$ counting the number of unmatched vertices among the available ones. The first-stage problem consists in finding
\begin{equation}
\label{two-stage}
  \bx_1^* = \argmin_{\bx_1} \mathbbm E_{\bt} \min_{\bx_2} \mathcal E(\bx_1, \bt, \bx_2)
\end{equation}
subject to the matching constraints
\begin{subnumcases}{\label{matching}}
  \sum_{l \in \partial r} x_{l r} \leq 1 & $(\forall r \in R)$ \label{matching_r}\\
  \sum_{r \in \partial l_1} x_{l_1 r} \leq 1 & $(\forall l_1 \in L_1)$ \label{matching_l1}\\
  \sum_{r \in \partial l_2} x_{l_2 r} \leq t_{l_2} & $(\forall l_2 \in L_2)$ \label{matching_l2}
\end{subnumcases}
where $\partial r = \{l \in L : (lr) \in E\}$ and similarly for $\partial l_1$ and $\partial l_2$. Once $\bf x_1$ and $\bf t$ are determined, it is straightforward to solve the second-stage problem. The difficulty of the problem stems from the fact that $\mathbbm E_\bt \min_{\bx_2} \mathcal E(\bx_1, \bx_2, \bt)$ has a highly non trivial dependence on $\bx_1$. In order to overcome this difficulty, we shall use the cavity method to first compute the minimum energy relative to $\bx_2$ for fixed $\bx_1$ and $\bt$, and then to compute the average over $\bt$ of this quantity.

In order to simplify the notation, in the following we shall \emph{always} assume (unless explicitly specified differently) that $l$ denotes a vertex in $L$, $l_1$ a vertex in $L_1$, $l_2$ a vertex in $L_2$ and $r$ a vertex in $r$, possibly restricted to the neighbors of some given node, and that $(lr)$ denotes an edge of the graph, $(l_1 r)$ an edge with $l_1 \in L_1$ and so on.

\subsection{Message passing solution of the second-stage problem}

Once $\bx_1$ is determined and the stochastic parameters $\bt$ are set, it is straightforward to find the optimal $\bx_2$. We shall now show how to do this is using MS, as discussed in \cite{Mezard-Zdeborova}. For each edge $(l_2 r) \in E$ we introduce the MS messages $m_{l_2 \to r}$ and $m_{r \to l_2}$. Notice that since the variables in the problem are defined on the edges of the original graph $G$, while the clauses are defined on its vertices, there is no distinction between ``clause to variable'' and ``variable to clause'' messages. The MS equations are then
\begin{align}
  m_{r \to l_2} &=
	\begin{dcases}
	  -\max[-1, \, \max_{l'_2 \in \partial r \sm l_2} m_{l'_2 \to r}] & \text{if $x_{l_1} = 0$ for each $l_1 \in \dd r$} \\
	  -\infty & \text{otherwise}
	\end{dcases} \label{MS0_r_to_l} \\
  m_{l_2 \to r} &=
	\begin{dcases}
	  -\max[-1, \, \max_{r' \in \partial l_2 \sm r} m_{r' \to l_2}] & \text{if $t_{l_2} = 1$} \\
	  -\infty & \text{otherwise}
	\end{dcases} \label{MS0_l_to_r}
\end{align}
where the condition in (\ref{MS0_r_to_l}) derives from the matching constraint (\ref{matching_r}) and the condition in (\ref{MS0_l_to_r}) derives from the matching constraint (\ref{matching_l2}).
Notice that the presence of $\max[-1, \cdots]$ in these equations makes it possible to replace the $-\infty$ with $-1$. This in turns allows to make a very useful simplification of the notation: we shall define the messages $m$ also on the edges connected to the vertices in $L_1$, with the convention that $m_{l_1 \to r} = m_{r \to l_1} = 1$ if $x_{l_1 r} = 1$ and $m_{l_1 \to r} = m_{r \to l_1} = -1$ if $x_{l_1 r} = 0$. It is easy to see that the previous equations then become
\begin{align}
  m_{r \to l_2} &= -\max[-1, \, \max_{l' \in \partial r \sm l2} m_{l' \to r}] \label{MS_r_to_l} \\
  m_{l_2 \to r} &=
	\begin{dcases}
	  -\max[-1, \, \max_{r' \in \partial l_2 \sm r} m_{r' \to l_2}] & \text{if $t_{l_2} = 1$} \\
	  -1 & \text{otherwise.}
	\end{dcases} \label{MS_l_to_r}
\end{align}

These equations can be solved by iteration, and knowing the value of the messages at the fixed point allows to compute
\begin{align}
\label{E_x1_t}
  \mathcal E^*(\bx_1, \bt) &= \min_{\bx_2} \mathcal E(\bx_1, \bt, \bx_2) \nonumber \\
  &= - \sum_{l} \max[-1,\, \max_{r \in \partial l} m_{r \to l}] - \sum_{r}\max[-1,\, \max_{l \in \partial r} m_{l \to r}] + \sum_{(l r)} \max[0,\, m_{l \to r} + m_{r \to l}] \, .
\end{align}
This expression is obtained by taking the zero-temperature limit of the first line of (21) in \cite{Mezard-Zdeborova} (notice that the simplified expression in the second line cannot be used in the zero-temperature limit, because it depends on an unresolved indetermination; this can be verified easily on a star-shaped or on a linear chain graph).
The dependence of this expression on $\bt$ is not explicit, and it derives from the matching constraints (\ref{matching_l2}) through the update equations (\ref{MS_l_to_r}).

For the sake of our computation, it is important to analyse the nature and the number of the fixed points of the MS equations (\ref{MS_r_to_l}, \ref{MS_l_to_r}). It is easily seen that these equations are closed for messages with support in $\{-1,1\}$, and also for messages with support in $\{-1,0,1\}$, so we can expect the fixed points to have support on either one of these sets. Fixed points with other support can exist for finite-size instances with appropriate initial values of the messages, but we have verified numerically that they disappear in the infinite-size limit, and we shall ignore them. A detailed analysis of the fixed points obtained in the infinite size limit is carried out in \cite{Mezard-Zdeborova} for the case of random graphs with average connectivity $c$, where it is shown that the fixed points with support in $\{-1,1\}$ (which we shall refer to as ``two-valued'' fixed points) correspond to replica-symmetric states and are correct for $c < e$, while the fixed points with support in $\{-1,0,1\}$ (which we shall refer to as ``three-valued'' fixed points) correspond to replica-symmetry-breaking states and are correct for $c > e$. In the remainder of this Paragraph we shall extend that analysis to the case of \emph{bipartite} random graphs, which is of interest for us.

Let us consider a uniform ensemble of instances with poissonian degree distribution and average degree $c$ (we shall consider balanced bipartite graphs for simplicity, so that the connectivity of left-hand nodes and that of right-hand nodes coincide), in the infinite-size limit. We shall denote by $P_{L \to R}^+$ the average fraction of messages $m_{l \to r}$ that are equal to $+1$ and by $P_{L \to R}^-$ the average fraction of messages $m_{l \to r}$ that are equal to $-1$, and similarly define $P_{R \to L}^+$ and $P_{R \to L}^-$. From their definitions and from the MS equations (\ref{MS_r_to_l}, \ref{MS_l_to_r}) one obtains that these quantities must satisfy the following equations:
\begin{align}
  P_{L \to R}^+ &= \sum_k e^{-c} \frac {c^k} {k!} \left(P_{R \to L}^-\right)^k = e^{-c (1 - P_{R \to L}^-)} \label{plrp} \\
  P_{L \to R}^- &= \sum_k e^{-c} \frac {c^k} {k!} \left[ 1 - \left( 1 - P_{R \to L}^+ \right)^k \right] = 1 - e^{-c P_{R \to L}^+} \label{plrm} \\
  P_{R \to L}^+ &= e^{-c(1 - P_{L \to R}^-)} \label{prlp} \\
  P_{R \to L}^- &= 1 - e^{-c P_{L \to R}^+} \label{prlm}
\end{align}
which implies that each of the quantities $P_{L \to R}^+$, $(1 - P_{L \to R}^-)$, $P_{R \to L}^+$ and $(1 - P_{R \to L}^-)$ must satisfy the equation
\begin{align}
  \label{expexp}
  x = \exp \left[ -c \exp(-c x) \right] \, .
\end{align}
The crucial difference between the case we consider and the non-bipartite case considered in \cite{Mezard-Zdeborova} is that here $P_{L \to R}^\pm$ can be different from $P_{R \to L}^\pm$ if (\ref{expexp}) admits more than one solution. On the other hand, it is always possible to find a solution with $P_{L \to R}^+ = 1 - P_{L \to R}^-$ (and then $P_{R \to L}^+ = 1 - P_{R \to L}^-$), so we expect that the two-valued fixed point is always present.

In fact, for $c < e$ (\ref{expexp}) admits a unique solution, and the distribution of the cavity messages will be unique and satisfy $P_{L \to R}^+ = 1 - P_{L \to R}^-$ and $P_{R \to L}^+ = 1 - P_{R \to L}^-$. This will correspond to an essentially unique fixed point of the MS equations: it is possible that some (small) isolated components admit several fixed points, multiplying the total number of fixed points, but the extensive component (which dominates the energy) will have a unique fixed point, and this will be a two-valued fixed point (i.e. with support in $\{-1, 1\}$). This statement is confirmed by numerical simulations.

For $c > e$ the situation is more complicated: (\ref{expexp}) admits 3 solutions, and $P_{L \to R}^+$ and $P_{R \to L}^-$ can be different from $1 - P_{L \to R}^-$ and $1 - P_{R \to L}^-$ respectively. From (\ref{prlp}, \ref{prlm}) we see that $P_{R \to L}^+$ and $P_{R \to L}^-$ are determined from $P_{L \to R}^+$ and $P_{L \to R}^-$, so the total number of solutions will depend on the number of solutions for $P_{L \to R}^+$ and for $1 - P_{L \to R}^-$ only. Taking into account the constraint $P_{L \to R}^+ \leq 1 - P_{L \to R}^-$, we see that the total number of solutions for the distribution of the cavity messages is at most 6. Some of these solutions however might correspond to negative values of the energy, and must therefore be rejected. As an example we have studied in detail the case $c = 5$, where the number of solutions with positive energy is 3, and they all have \emph{exactly} the same energy. One of the 3 solutions is three-valued, and the remaining 2 are two-valued.
On finite size instances, we have verified numerically that these 3 fixed points can always be obtained by chosing appropriate initial conditions. Their energies are close to each other, but not exactly the same, and the correct one is always the largest.

We conclude from this discussion that the energy computed from (\ref{E_x1_t}) is correct for instances extracted with poissonian degree distributions with $c < e$ and approximately correct for instances with $c > e$. It must be noted, however, that the reduced instance to be solved in the second stage is not necessarily poissonian, as the probability that a node in $R$ is matched to a node in $L_1$ can be correlated to its degree. Moreover, it is possible that some small disconnected components have multiple solutions, that combined with the 3 solutions of the giant component give a larger number of fixed points, but these will always have energies that are approximately equal. We shall neglect these possible issues, comforted by our numerical results.

In the following, we shall treat the two- and three-valued cases separately: we shall see that they give rise to different algorithms for the optimization over $\bx_1$. Based on the above discussion, we expect that the two-valued algorithm will find the correct solution for poissonian instances with small connectivity, while the three-valued algorithm will do it for instances with large connectivity. We can also expect the two-valued algorithm to provide an approximate solution for large connectivity, and in fact we shall see that the average energy it obtains on the random ensemble of instances we have analysed numerically is almost exactly the same as that obtained by the three-valued algorithm.

\subsection{Message passing solution of the first-stage problem in the two-valued case}

\subsubsection{Computing the average over $\bt$}

We shall now compute the average over $\bt$ of the expression (\ref{E_x1_t}),
\begin{align}
  \mathcal E^*(\bx_1) = \mathbbm E_{\bt} \min_{\bx_2} \mathcal E(\bx_1, \bt, \bx_2)
\end{align}
for the two-valued case where the MS messages $m_{l_2 \to r}$ and $m_{r \to l_2}$ take values in $\{-1,1\}$. For this purpose, we shall treat the quantities $m_{l_2 \to r}$, $m_{r \to l_2}$ and $t_{l_2}$ as random variables with a joint probability distribution
\begin{align}
\label{Q(m,t)}
  \mathcal Q(\bM, \bt) \propto \prod_{l_2} p_{l_2}(t_{l_2}) \times \prod_{(l_2 r)} \mathbbm 1 \left[ m_{l_2 \to r} = \hat m_{l_2 \to r} \right] \mathbbm 1 \left[ m_{r \to l_2} = \hat m_{r \to l_2} \right]
\end{align}
where $\hat m_{l_2 \to r}$ is a shorthand for $\hat m_{l_2 \to r}( \{m_{r' \to l_2}, \, r' \in \dd l_2 \sm r\}, t_{l_2} )$ defined as the right-hand side of (\ref{MS_l_to_r}) and similarly $\hat m_{r \to l_2}$ is a shorthand for $\hat m_{r \to l_2}(\{m_{l'_2 \to r}, \, l'_2 \in \dd r \sm l_2 \})$ defined as the right-hand side of (\ref{MS_r_to_l}). We then need to compute the average of (\ref{E_x1_t}) relative to this distribution.

Within the cavity approximation, we follow the approach of Section \ref{sec:general_approach} and introduce the cavity marginals $P_{l_2 \to r}(m_{l_2 \to r})$ and $P_{r \to l_2}(m_{r \to l_2})$. Since $m_{l_2 \to r}$ and $m_{r \to l_2}$ have support in $\{-1, 1\}$, we can parametrize these marginals with a single real number, $P_{l_2 \to r} = \pr_{l_2 \to r}[m_{l_2 \to r} = 1] \in [0,1]$ (and similarly for $P_{r \to l_2}$). The update equations for these cavity marginals can be obtained with the general method of the previous section (i.e. using BP for the distribution $\mathcal Q(\bM, \bt)$), but in this case it is possible to derive them in a more intuitive way as follows. Again, we can simplify the notation by extending the definition of $P_{l \to r}$ and $P_{r \to l}$ also to the edges connected to vertices in $L_1$, with the convention that $P_{l_1 \to r} = P_{r \to l_1} = 1$ if $x_{l_1} = 1$, and $P_{l_1 \to r} = P_{r \to l_1} = 0$ if $x_{l_1} = 0$.

We see from (\ref{MS_r_to_l}) that $m_{r \to l_2}$ is $+1$ if and only if all the incoming $m_{l' \to r}$ are $-1$ (for each $l' \in \dd r \sm l_2$), so that
\begin{align}
  \label{MS2_r_to_l}
  P_{r \to l_2} = \prod_{l' \in \dd r \sm l_2} (1 - P_{l' \to r}) \, .
\end{align}
Similarly, from (\ref{MS_l_to_r}) we see that $m_{l_2 \to r}$ is $+1$ if and only if $t_{l_2}$ is 1 (which happens with probability $p_{l_2}$), and all the incoming $m_{r' \to l_2}$ are $-1$ (for each $r' \in \dd l_2 \sm r$), so that
\begin{align}
  \label{MS2_l_to_r}
  P_{l_2 \to r} = p_{l_2} \prod_{r' \in \dd l_2 \sm r} (1 - P_{r' \to l_2}) \, .
\end{align}
Notice that by doing this (i.e. conditioning the probabilities to the values of $t_{l_2}$) we are giving the correct weight to all the fixed points even in the case where their number varies with $\bt$, as explained in Section \ref{sec:general_approach}.

The coupled equations (\ref{MS2_r_to_l}, \ref{MS2_l_to_r}) can again be solved by iteration, and $\mathcal E^*(\bx_1)$ can be computed from the fixed point values of $P_{l \to r}$ and $P_{r \to l}$.
The contribution of a vertex $l \in L$ will be different from zero only if the node is available for matching, which happens with probability $p_l$ (setting $p_l = 1$ if $l \in L_1$), and in this case it will be equal to $-\max[-1, \max_{r \in \dd l} m_{r \to l}]$ which is $+1$ with probability $\prod_{r \in \dd l} (1 - P_{r \to l})$, and $-1$ otherwise, so that
\begin{align}
  \mathbbm E_{\bt} \left[ - \max[-1, \max_{r \in \dd l} m_{r \to l}] \right] = p_l \left[ 2 \prod_{r \in \dd l}(1 - P_{r \to l}) - 1 \right] \, .
\end{align}
Similarly, the contribution of a vertex $r \in R$ is
\begin{align}
  \mathbbm E_{\bt} \left[ - \max[-1, \max_{l \in \dd r} m_{l \to r}] \right] = \left[ 2 \prod_{l \in \dd r}(1 - P_{l \to r}) - 1 \right] \, .
\end{align}
Finally, the contribution from an edge $(lr) \in E$ in (\ref{E_x1_t}) is $\max[0, \, m_{l \to r} + m_{r \to l}]$ which is $+2$ with probability $P_{l \to r} \times P_{r \to l}$ and zero otherwise, so that
\begin{align}
  \mathbbm E_{\bt} \big[ \max[0, \, m_{l \to r} + m_{r \to l}] \big] = 2 P_{l \to r} P_{r \to l} \, .
\end{align}
We obtain
\begin{align}
\label{E_x1}
  \mathcal E^*(\bx_1) &= \sum_l p_l \left[ 2 \prod_{r \in \dd l}(1 - P_{r \to l}) - 1 \right] + \sum_r \left[ 2 \prod_{l \in \dd r}(1 - P_{l \to r}) - 1 \right] + 2 \sum_{(lr)} P_{l \to r} P_{r \to l} \, .
\end{align}
Let us stress again that in this expression, even though $L_1$ and $L_2$ vertices are treated in a completely symmetric way, the values of $P_{l_1 \to r}$ and $P_{r \to l_1}$ with $l_1 \in L_1$ are explicitly determined by $\bx_1$, while the values of $P_{l_2 \to r}$ and $P_{r \to l_2}$ with $l_2 \in L_2$ are determined by $\bx_1$ \emph{implicitly} through the update equations (\ref{MS2_r_to_l}, \ref{MS2_l_to_r}).

\subsubsection{Solving for $\bx_1$}

We are now in position to solve the first stage problem, namely finding $\bx_1^*$ defined as
\begin{align}
  \bx_1^* = \argmin_{\bx_1} \mathbbm E_{\bt} \min_{\bx_2} \mathcal E(\bx_1, \bt, \bx_2)
	= \argmin_{\bx_1} \mathcal E^*(\bx_1)
\end{align}
where $\mathcal E^*(\bx_1)$ is defined in (\ref{E_x1}), and where we remind that $P_{l_1 \to r} = P_{r \to l_1} = x_{l_1 r}$, with $x_{l_1 r}$ subject to the matching constraints (\ref{matching_r}, \ref{matching_l1}), while  $P_{l_2 \to r}$ and $P_{r \to l_2}$ must satisfy the update equations (\ref{MS2_r_to_l}, \ref{MS2_l_to_r}), and we keep following the convention that $l_1$ always denotes a vertex in $L_1$ and $l_2$ a vertex in $L_2$, and similarly for $l_1', l_2'$ etc.

Since $\mathcal E^*(\bx_1)$ is a sum of local terms subject to local constraints, we can solve this minimization problem with MS. We introduce the messages in terms of the cavity marginals $\pr_{l \to r}[P_{l \to r} = P]$ and $\pr_{r \to l}[P_{r \to l} = P]$. For the edges connected to vertices in $L_1$ we define
\begin{align}
  \Psi_{l_1 \to r} &= \log \frac {\pr_{l_1 \to r}[ P_{l_1 \to r} = 1]} {\pr_{l_1 \to r}[ P_{l_1 \to r} = 0]} = \log \frac {\pr_{l_1 \to r}[ x_{l_1 r} = 1]} {\pr_{l_1 \to r}[ x_{l_1 r} = 0]} \, , \\
  \Psi_{r \to l_1} &= \log \frac {\pr_{r \to l_1}[ P_{r \to l_1} = 1]} {\pr_{r \to l_1}[ P_{r \to l_1} = 0]} = \log \frac {\pr_{r \to l_1}[ x_{l_1 r} = 1]} {\pr_{r \to l_1}[ x_{l_1 r} = 0]}
\end{align}
which are simply real numbers. On the other hand, for the edges connected to vertices in $L_2$ the probabilities $P_{l_2 \to r}$ and $P_{r \to l_2}$ can take any real value in $[0, 1]$ and the messages, defined as
\begin{align}
  \Psi_{l_2 \to r}(P) &= \log \pr_{l_2 \to r}[ P_{l_2 \to r} = P ] + C_{l_2 \to r} \, , \\
  \Psi_{r \to l_2}(P) &= \log \pr_{r \to l_2}[ P_{r \to l_2} = P ] + C_{r \to l_2}
\end{align}
are functions of a real variable. The additive constants $C_{l_2 \to r}$ and $C_{r \to l_2}$ are set by requiring that $\max_P \Psi_{l_2 \to r}(P) = \max_P \Psi_{r \to l_2}(P) = 0$. For numerical purposes these functions can be approximated by a vector of real negative numbers corresponding to finite size bins for the values of $P$ in $[0, 1]$.

In order to obtain the update equations for $\Psi_{l \to r}$ (and $\Psi_{r \to l}$), we must consider all the terms in (\ref{E_x1}) where the corresponding variable $P_{l \to r}$ (and $P_{r \to l}$) appears, which will always be two: a vertex term and an edge term. Notice that the fact that we include the edge term in both the updates of $\Psi_{l \to r}$ and $\Psi_{r \to l}$ implies that we are defining these messages as the ``variable to clause'' ones. It will be important to remember this when deciding the value of $x_{l_1 r}$ from the values of the fixed point messages $\Psi_{l_1 \to r}$ and $\Psi_{r \to l_1}$, since we shall have to subtract from $\Psi_{l_1 \to r} + \Psi_{r \to l_1}$ the edge contribution to the energy to avoid double-counting it.

The update equations are obtained in a straightforward manner as follows.
Let us begin with the equation for $\Psi_{l_1 \to r}$. The relevant energy contribution is
\begin{align}
  \mathcal E_{l_1 \to r}(P_{l_1 \to r}, \{ P_{r' \to l_1} \, , r' \in \dd l_1 \sm r \}) = \left[ 2 \prod_{r' \in \dd l_1} ( 1 - P_{r' \to l_1} ) - 1 \right] + 2 P_{l_1 \to r} P_{r \to l_1}
\end{align}
where we recall that $P_{l_1 \to r'} = P_{r' \to l_1} = x_{l_1 r'} \in \{0, 1\}$ for each $r' \in \dd l_1$, and that these are subject to the matching constraint (\ref{matching_l1}). We obtain
\begin{align}
  \log \pr_{l_1 \to r} [ P_{l_1 \to r} = P ] = \max_{
	\substack{
	  \{ P_{r' \to l_1} \in \{0,1\},\, r' \in \dd l_1 \sm r\} \text{ s.t:} \\
	  P + \sum_{r' \in \dd l_1 \sm r} P_{r' \to l_1} \leq 1
	}
  }\left[ -\mathcal E_{l_1 \to r}(P, \{ P_{r' \to l_1}, r' \in \dd l_1 \sm r \}) + \sum_{r' \in \dd l_1 \sm r} \log \pr_{r' \to l}[P_{r' \to l}] \right] \, .
\end{align}
When $P_{l_1 \to r}$ is 1, the matching constraint (\ref{matching_l1}) forces all the incoming $P_{r' \to l_1}$ to be 0 (for $r' \in \dd l_1 \sm r$), and the previous equation reduces to
\begin{align}
\label{MS3_l1_to_r_1}
  \log \pr_{l_1 \to r} [ P_{l_1 \to r} = 1 ] = - 1 + \sum_{r' \in \dd l_1 \sm r} \log \pr_{r' \to l_1}[ P_{r' \to l_1} = 0 ] \, .
\end{align}
When $P_{l_1 \to r}$ is 0, because of the matching constraint (\ref{matching_l1}) the incoming $P_{r' \to l_1}$ (with $r' \in \dd l_1 \sm r$) can either be all 0 (as before), or one of them can be equal to 1, all the other ones being 0. We find
\begin{align}
\label{MS3_l1_to_r_0}
  \log \pr_{l_1 \to r}[P_{l_1 \to r} = 0 ] = \max \Bigg\{ & -1 + \sum_{r' \in \dd l_1 \sm r} \log \pr_{r' \to l_1}[ P_{r' \to l_1} = 0 ] \, , \nonumber \\
  & \max_{r' \in \dd l_1 \sm r} \Bigg[ 1 + \log P_{r' \to l_1}[ P_{r' \to l_1} = 1 ] + \sum_{r'' \in \dd l_1 \sm \{r, r'\}} \log \pr_{r'' \to l_1} [ P_{r'' \to l_1} = 0 ] \Bigg] \Bigg\} \, .
\end{align}
By subtracting (\ref{MS3_l1_to_r_0}) from (\ref{MS3_l1_to_r_1}) we obtain
\begin{align}
  \label{Psi_l1_to_r}
  \Psi_{l_1 \to r} = - \max \left[ 0, \, 2 + \max_{r' \in \dd l_1 \sm r} \Psi_{r' \to l_1} \right] \, .
\end{align}

In a similar way we obtain the update equations for $\Psi_{l_2 \to r}(P_{l_2 \to r})$. The main difference is that now both $P_{l_2 \to r}$ and the variables associated to the incoming messages $\Psi_{r' \to l_2}(P_{r' \to l_2})$ will be \emph{real} variables in the interval $[0, 1]$. The maximisation is therefore no longer performed over a discrete set of partial configurations for the incoming variables, but on a continuous range. Moreover, these variables have to satisfy the update equations (\ref{MS2_r_to_l}, \ref{MS2_l_to_r}), which are a constraint in the maximisation. The relevant energy contribution is again formed by a vertex and an edge term and is given by
\begin{align}
  \mathcal E_{l_2 \to r}(P_{l_2 \to r}, \{ P_{r' \to l_2}, \, r' \in \dd l_2 \sm r \}) &= p_{l_2} \left[ 2 \prod_{ r' \in \dd l_2 } ( 1 - P_{r' \to l_2} ) - 1 \right] + 2 P_{l_2 \to r} P_{r \to l_2} \\
  &= \left[ 2 (1 - P_{r \to l_2}) P_{l_2 \to r} - p_{l_2} \right] + 2 P_{l_2 \to r} P_{r \to l_2} \\
  &= 2 P_{l_2 \to r} - p_{l_2}
\end{align}
where we have made use of the update equation (\ref{MS2_l_to_r}). Notice that, as expected from the discussion in Section \ref{sec:general_approach}, this energy term \emph{does not} depend on $P_{r \to l_2}$, thanks to the simplification between part of the vertex contribution and the edge contribution. This is crucial to ensure that the messages $\Psi_{l_2 \to r}(P)$ and $\Psi_{r \to l_2}(P)$ are uncorrelated, as required in order to apply the cavity method. In fact, in this case the result is extraordinarily simple, as it only depends on the outgoing variable $P_{l_2 \to r}$ and is independent of the incoming variables $\{ P_{r' \to l_2}, \, r' \in \dd l_2 \sm r \}$.
The update equation is then obtained in a straightforward manner:
\begin{align}
  \Psi_{l_2 \to r}(P) &= \max_{\substack{
	\{ P_{r' \to l_2} \in [0,1], \, r' \in \dd l_2 \sm r \} \text{ s.t:} \\
	P = p_{l_2} \prod_{r' \in \dd l_2 \sm r} (1 - P_{r' \to l_2})
  }} \left[ - \mathcal E_{l_2 \to r}(P) + \sum_{r' \in \dd l_2 \sm r} \Psi_{r' \to l_2}(P_{r' \to l_2}) \right] \\
  &= -(2 P - p_{l_2}) +  \max_{\substack{
	\{ P_{r' \to l_2} \in [0,1], \, r' \in \dd l_2 \sm r \} \text{ s.t:} \\
	P = p_{l_2} \prod_{r' \in \dd l_2 \sm r} (1 - P_{r' \to l_2})
  }} \left[ \sum_{r' \in \dd l_2 \sm r} \Psi_{r' \to l_2}(P_{r' \to l_2}) \right] \, . \label{Psi_l2_to_r}
\end{align}

It is important to realize that this constrained maximisation problem can be solved efficiently by exploiting the associativity of the maximum as follows. In order to see how, let us introduce the following simplified notation:
\begin{align}
  g_k(y) &= \max_{ \substack{
	\{z_1, \dots, z_k\} \in [0,1]^k \text{ s.t.: } \\
	y \, = \, \prod_{j=1, \dots, k} (1-z_k)
  }} \sum_{j=1, \dots, k} f_j(z_j)
\end{align}
with $y \in [0,1]$, which is of the form we need. Then
\begin{align}
  g_k(y) &= \max_{z_k \in [0, 1-y]} \left[ f_k(z_k) + \max_{ \substack{
	\{z_1, \dots, z_{k-1} \} \in [0,1]^{k-1} \text{ s.t.: } \\
	y/(1-z_k) \, = \, \prod_{j=1, \dots, k-1} (1-z_j)
  }} \sum_{j=1, \dots, k-1} f_j(z_j) \right] \\
  &= \max_{z_k \in [0, 1-y]} \left[ f_k(z_k) + g_{k-1} \left( \frac y {1-z_k} \right) \right] \label{efficient_max}
\end{align}
so we can compute $g_k(y)$ iteratively, starting with $g_0(y) = 0$, in a time which is linear in the number of variables appearing in the maximum (i.e. linear in the connectivity of $l_2$). In fact, it is possible to compute \emph{all} the messages $\Psi_{l_2 \to r}(P)$ for $r \in \dd l_2$ in a time which is linear in the connectivity by computing (and keeping in memory) all the quantities (resuming to the full notation)
\begin{align}
  g_r(y) &= \max_{ \substack{
	\{P_{r' \to l_2} \in [0,1], \, r' \in \dd l_2 :\, r' < r \} \text{ s.t.: } \\
	y \, = \, \prod_{r' \in \dd l_2 :\, r' < r} (1 - P_{r' \to l_2})
  }} \left[ \sum_{r' \in \dd l_2 :\, r' < r} \Psi_{r' \to l_2}(P_{r' \to l_2}) \right] \, ,\\
  h_r(y) &= \max_{ \substack{
	\{P_{r' \to l_2} \in [0,1], \, r' \in \dd l_2 :\, r' > r \} \text{ s.t.: } \\
	y \, = \prod_{r' \in \dd l_2 :\, r' > r} (1 - P_{r' \to l_2})
  }} \left[ \sum_{r' \in \dd l_2 :\, r' > r} \Psi_{r' \to l_2}(P_{r' \to l_2}) \right]
\end{align}
which are computed iteratively as in (\ref{efficient_max}) and in terms of which the update equation (\ref{Psi_l2_to_r}) becomes
\begin{align}
  \Psi_{l_2 \to r}(P) &= - (2 P - p_{l_2}) + \max_{\substack{
	\{ y_1, y_2 \} \in [0,1]^2 \text{ s.t.: } \\
	P = p_{l_2} y_1 y_2
  }} \left[ g_r(y_1) + h_r(y_2) \right] \, .
\end{align}
By doing this, all the outgoing messages are computed by performing less than $3|\dd l_2|$ one-dimensional maximisations, each of which has a time complexity proportial to the $B^2$, the square of the number of bins that represent the cavity marginals. Overall the time complexity is proportional to $B^2 |E|$ per iteration.

Let us now turn to the update equation for $\Psi_{r \to l_1}$. In this case, the variable $P_{r \to l_1}$ on the output edge is in $\{0, 1\}$ and the corresponding message is a real number, while the variables and messages on the incoming edges $(lr)$ with $l \in \dd r \sm l_1$ are mixed: if $l \in L_1$, the variable $P_{l \to r}$ will be in $\{0, 1\}$ and the corresponding message will be a real number, and if $l \in L_2$ the variable $P_{l \to r}$ will be in $[0, 1]$ and the corresponding message will be a function with domain in $[0,1]$ and codomain in $]-\infty, 0]$. The integer variables $P_{l'_1 \to r}$ must satisfy the matching constraints (\ref{matching_r}), that is $P_{r \to l_1} + \sum_{l'_1 \in \dd r \sm l_1} P_{l'_1 \to r} \leq 1$. On the other hand, the constraint that the real variables $P_{l_2 \to r}$ must satisfy the update equations (\ref{MS2_r_to_l}) has no effect, since the values of $P_{r \to l_2}$ for each $l_2 \in \dd r \cap L_2$ do not appear in the update. The relevant energy contribution is as usual formed by a vertex term and an edge term and is equal to
\begin{align}
  \mathcal E_{r \to l_1}(P_{r \to l_1}, \{ P_{l' \to r }, \, l' \in \dd r \sm l_1 \} ) &= 2 \prod_{l' \in \dd r} ( 1 - P_{l' \to r} ) - 1 + 2 P_{l_1 \to r} P_{r \to l_1}
\end{align}
where we recall that $P_{l_1 \to r} = P_{r \to l_1} = x_{l_1 r}$. The cavity marginals are then
\begin{align}
  \label{Psi_r_to_l1_1}
  \log \pr_{r \to l_1}[ P_{r \to l_1} = 1 ] &= \max_{\{ P_{l_2 \to r}, \, l_2 \in \dd r\}} \Bigg\{ -1 + \sum_{l'_1 \in \dd r \sm l_1} \log \pr_{l'_1 \to r}[ P_{l'_1 \to r} = 0 ] + \sum_{l_2 \in \dd r} \Psi_{l_2 \to r} ( P_{l_2 \to r} ) \Bigg\} \, , \\
  \label{Psi_r_to_l1_0}
  \log \pr_{r \to l_1}[ P_{r \to l_1} = 0 ] &= \max \Big[ A \big( \{ \Psi_{l \to r}, l \in \dd r \sm l_1 \} \big), B  \big( \{ \Psi_{l \to r}, l \in \dd r \sm l_1 \} \big) \Big]
\end{align}
with
\begin{multline}
  A \big( \{ \Psi_{l \to r}, l \in \dd r \sm l_1 \} \big) \\
	= \max_{\{ P_{l_2 \to r}, \, l_2 \in \dd r\}} \Bigg[ 1 - 2 \prod_{l_2 \in \dd r} (1 - P_{l_2 \to r}) + \sum_{l'_1 \in \dd r \sm l_1} \log \pr_{l'_1 \to r}[ P_{l'_1 \to r} = 0 ] + \sum_{l_2 \in \dd r} \Psi_{l_2 \to r} ( P_{l_2 \to r} ) \Bigg] \, ,
\end{multline}
and
\begin{multline}
  B \big( \{ \Psi_{l \to r}, l \in \dd r \sm l_1 \} \big) \\
	= \max_{\{ P_{l_2 \to r}, \, l_2 \in \dd r\}} \Bigg[ 1 + \max_{l'_1 \in \dd r \sm l_1} \Bigg( \log \pr_{l'_1 \to r} [ P_{l'_1 \to r} = 1 ] + \sum_{l''_1 \in \dd r \sm \{l_1, l'_1\}} \log \pr_{l''_1 \to r}[ P_{l''_1 \to r} = 0 ] \Bigg) + \\
	+ \sum_{l_2 \in \dd r} \Psi_{l_2 \to r}(P_{l_2 \to r}) \Bigg]\, .
\end{multline}
Notice that all the maximisations in (\ref{Psi_r_to_l1_1}, \ref{Psi_r_to_l1_0}) are \emph{unconstrained} and therefore performed easily. The outgoing message $\Psi_{r \to l_1}$ is then computed by subtracting (\ref{Psi_r_to_l1_0}) from (\ref{Psi_r_to_l1_1}).

Finally, we turn to the update equation for $\Psi_{r \to l_2}(P_{r \to l_2})$. Exactly as above, the incoming variables and messages will be of mixed types. The only difference is that now the message to be computed is a function, and that the incoming variables $P_{l' \to r}$ with $l' \in \dd r \sm l_2$ must satisfy the update equation (\ref{MS2_r_to_l}) for $P_{r \to l_2}$. As before, the relevant energy contribution can be simplified by using this constraint to eliminate the dependence on $P_{l_2 \to r}$, whose presence would undermine the application of the cavity method:
\begin{align}
  \mathcal E_{r \to l_2} (P_{r \to l_2}, \{ P_{l' \to r}, \, l' \in \dd r \sm l_2 \}) &= \left[ 2 \prod_{l' \in \dd r} (1 - P_{l' \to r}) - 1 \right] + 2 P_{r \to l_2} P_{l_2 \to r} \\
  &= 2(1 - P_{l_2 \to r}) P_{r \to l_2} - 1 + 2 P_{l_2 \to r} P_{r \to l_2} \\
  &= 2 P_{r \to l_2} - 1 \, .
\end{align}
We then have:
\begin{align}
  \log \pr_{r \to l_2}[ P_{r \to l_2} = P ] = \max_{\substack{
	\{ P_{l'_2 \to r} \in [0,1], \, l'_2 \in (\dd r \sm l_2) \cap L_2 \} \, , \\
	\{ P_{l_1 \to r} \in \{0,1\}, \, l_1 \in \dd r \cap L_1 \} \text{ s.t.: } \\
	\sum_{l_1 \in \dd r \cap L_1} P_{l_1 \to r} \leq 1 \, ,\\
	P = \prod_{l'_2 \in \dd r \sm l_2} ( 1 - P_{l'_2 \to r} )
  }} \left[ 1 - 2 P + \sum_{l \in \dd r \sm l_2} \log \pr_{l \to r}[ P_{l \to r} ] \right]
\end{align}
which again can be computed efficiently thanks to (\ref{efficient_max}). It is straightforward to substitute the messages $\Psi_{r \to l_2}$ and $\Psi_{l \to r}$ instead of the cavity marginals.

The coupled update equations for $\Psi_{l_1 \to r}$, $\Psi_{r \to l_1}$, $\Psi_{l_2 \to r}$ and $\Psi_{r \to l_2}$ are solved by iteration. Notice that these are \emph{the only} message passing equations that actually have to be implemented and solved numerically. The message passing equations for the second-stage MS messages $m_{l \to r}$ and $m_{r \to l}$, as well as those for the messages $P_{l \to r}$ and $P_{r \to l}$ introduced for the computation of the average over $\bt$, are only needed to derive the expression of the constraints to which the maximisations in the update equations for $\Psi_{l_1 \to r}$, $\Psi_{r \to l_1}$, $\Psi_{l_2 \to r}$ and $\Psi_{r \to l_2}$ are subject to.

The optimal configuration $\bx_1$ is determined from the fixed-point value of the messages $\Psi_{l_1 \to r}$ and $\Psi_{r \to l_1}$ as
\begin{align}
  x_{l_1 r} =
	\begin{cases}
	  1 & \text{ if } \Psi_{l_1 \to r} + \Psi_{r \to l_1} + 2 > 0 \\
	  0 & \text{ otherwise}
	\end{cases}
\end{align}
The constant addend $+2$ in the condition is due to the fact that the energy term $2 P_{l_1 \to r} P_{r \to l_1} = 2 x_{l_1 r}$ associated to the edge $(l_1 r)$ is subtracted from the updates of both $\Psi_{l_1 \to r}$ and $\Psi_{r \to l_1}$, so that it must be added back to the sum $\Psi_{l_1 \to r} + \Psi_{r \to l_1}$ in order to avoid double-counting it.

In order to improve the convergence of the update equations we added both a noise and a reinforcement term to the messages propagating on the edges incident on vertices $l_1 \in L_1$ (i.e. to the messages that are needed to assign the binary variables $x_{l_1 r}$), with a technique similar to the one used in \cite{braunstein2006learning, AdWords}. The noise term is just a constant (small) random field $\eta_{l_1 r}$ acting on each variable $x_{l_1 r}$. Each $\eta_{l_1 r}$ is extracted uniformly in $[0, \bar \eta]$. The reinforcement term is defined as follows. We introduce an esternal field $H_{l_1 r}$ acting on each variable $x_{l_1 r}$ and a constant parameter $\rho$. Let us denote by a superscript the time $t$ in the update sequence (i.e. the iteration number). The update equation (\ref{Psi_l1_to_r}) becomes
\begin{align}
  \Psi_{l_1 \to r}^t &= \hat \Psi_{l_1 \to r}( \{ \Psi_{r' \to l_1}^{t-1}, \, r' \in \dd l_1 \sm r \} ) + t \rho H_{l_1 r}^{t-1} + \eta_{l_1 r}
\end{align}
where $\hat \Psi_{l_1 \to r}( \{ \Psi_{r' \to l_1}^{t-1}, \, r' \in \dd l_1 \sm r \} )$ is the right-hand side of (\ref{Psi_l1_to_r}) computed at time $t-1$. The update equation for $\Psi_{r \to l_1}^t$ is modified exaclty in the same way by adding the term $t \rho H_{l_1 r}^{t-1} + \eta_{l_1 r}$. The external field is computed after each iteration $t$ as the \emph{total} field acting on the variable at the iteration $t$,
\begin{align}
  H_{l_1 r}^t &= \Psi_{l_1 \to r}^t + \Psi_{r \to l_1}^t - t \rho H_{l_1 r}^{t-1} - \eta_{l_1 r} + 2
\end{align}
where the terms $- t \rho H_{l_1 r}^{t-1} - \eta_{l_1 r} + 2$ are again included in the sum $\Psi_{l_1 \to r}^t + \Psi_{r \to l_1}^t$ in order to avoid double-counting them, and with $H_{l_1 r}^0 = 0$.
In the presence of a reinforcement term, the values of the messages do not converge to a fixed point: on the contrary, typically some of them diverge. The convergence criterion used to stop the algorithm is then that the configuration of $\bx_1$ variables corresponding to the instantaneous value of the messages does not change for a number of iterations $I$ (e.g. $I = 100$).

\subsection{Message passing solution of the first-stage problem in the three-valued case}

Let us now turn to the case in which the MS messages $m_{l_2 \to r}$ and $m_{r \to l_2}$ introduced for the optimization over $\bx_2$ take the three values $\{-1, 0, 1\}$. We shall see that this leads to different update equations for the second and third level messages ($P$ and $\Psi(P)$ respectively), and therefore to a different algorithm. We shall proceed in close analogy to the two-valued case just discussed.

\subsubsection{Computing the average over $\bt$}

As before, we start by computing the average over $\bt$ the expression (\ref{E_x1_t}),
\begin{align}
  \label{E*3_x1}
  \mathcal E^*(\bx_1) = \mathbbm E_{\bt} \min_{\bx_2} \mathcal E(\bx_1, \bt, \bx_2) \, .
\end{align}
The joint distribution of the messages $m_{l_2 \to r}$ and $m_{r \to l_2}$ and of the stochastic parameters $\bt$ has the same expression (\ref{Q(m,t)}) as before, with the same update equations (\ref{MS_l_to_r}) and (\ref{MS_r_to_l}) defining the functions $\hat m_{l \to r}$ and $\hat m_{r \to l}$ respectively.

As before we introduce the cavity marginals $P_{l_2 \to r}(m_{l_2 \to r})$ and $P_{r \to l_2}(m_{r \to l_2})$, but since now $m_{l_2 \to r}$ and $m_{r \to l_2}$ take values in $\{-1, 0, 1\}$, in order to parametrize them we need the three real numbers $P^+_{l_2 \to r} = \pr[m_{l_2 \to r} = +1]$, $P^0_{l_2 \to r} = \pr[m_{l_2 \to r} = 0]$ and $P^-_{l_2 \to r} = \pr[m_{l_2 \to r} = -1]$ subject to the normalization condition $P^+_{l_2 \to r} + P^0_{l_2 \to r} + P^-_{l_2 \to r} = 1$ (and similarly for $P^+_{r \to l_2}$, $P^0_{r \to l_2}$ and $P^-_{r \to l_2}$). In order to simplify the notation, we introduce similar quantities for the edges incident on the  vertices in $L_1$ with the definitions $P^+_{l_1 \to r} = P^+_{r \to l_1} = 1$ if $x_{l_1 r} = 1$ and $P^-_{l_1 \to r} = P^-_{r \to l_1} = 1$ if $x_{l_1 r} = 0$.

From the update equation (\ref{MS_r_to_l}) we see that $m_{r \to l}$ is $+1$ if and only if all the incoming messages are $-1$ (for each $l' \in \dd r \sm l$), so that
\begin{align}
  \label{P+_r_to_l}
  P^+_{r \to l} = \prod_{l' \in \dd r \sm l} P^-_{l' \to r} \, .
\end{align}
Moreover, $m_{r \to l}$ is $-1$ if and only if at least one of the incoming $m_{l' \to r}$ (with $l' \in \dd r \sm l$) is $+1$, so that
\begin{align}
  \label{P-_r_to_l}
  P^-_{r \to l} = 1 - \prod_{l' \in \dd r \sm l} \left( 1 - P^+_{l' \to r} \right) \, .
\end{align}

Similarly, we see from the update equation (\ref{MS_l_to_r}) that $m_{l \to r}$ is $+1$ if and only if $t_l = 1$ (which happens with probability $p_l$), and all the incoming messages are $-1$ (for each $r' \in \dd l \sm r)$, so that
\begin{align}
  \label{P+_l_to_r}
  P^+_{l \to r} = p_l \prod_{r' \in \dd l \sm r} P^-_{r' \to l}
\end{align}
while $m_{l \to r}$ is $-1$ if $t_l = 0$ (which happens with probability $1 - p_l$), or if $t_l = 1$ and at least one of the incoming messages $m_{r' \to l}$ (with $r' \in \dd l \sm r$) is $+1$, so that
\begin{align}
  \label{P-_l_to_r}
  P^-_{l \to r} = (1 - p_l) + p_l \left[ 1 - \prod_{r' \in \dd l \sm r} \left( 1 - P^+_{r' \to l} \right) \right] \, .
\end{align}

Solving these coupled equations by iteration, we can compute $\mathcal E^*(\bx_1)$ in (\ref{E*3_x1}) as a function of the messages $P^{\pm}_{l \to r}$ and $P^{\pm}_{r \to l}$. We start by noticing that when $m_{r \to l} \in \{-1, 0, 1\}$ we have $-\max[-1, \max_{r \in \dd l} m_{r \to l}] = -\max_{r \in \dd l} m_{r \to l}$, and that the probability that $\max_{r \in \dd l} m_{r \to l} = -1$ is $p_l \prod_{r \in \dd l} P^-_{r \to l}$, while the probability that $\max_{r \in \dd l} m_{r \to l} = +1$ is $p_l [ 1 - \prod_{r \in \dd l} (1 - P^+_{r \to l})]$, so that
\begin{align}
  \mathbbm E_\bt \left[ -\max [ -1, \max_{r \in \dd l} m_{r \to l} ] \right] &= p_l \left[ \prod_{r \in \dd l} P^-_{r \to l} + \prod_{r \in \dd l} \left( 1 - P^+_{r \to l} \right) - 1 \right] .
\end{align}
Similarly,
\begin{align}
  \mathbbm E_\bt \left[ -\max [-1, \max_{l \in \dd r} m_{l \to r}] \right] &= \prod_{l \in \dd r} P^-_{l \to r} + \prod_{l \in \dd r} \left( 1 - P^+_{l \to r} \right) - 1 \, .
\end{align}
Finally, $\max[0, m_{r \to l} + m_{l \to r}] = 2$ with probability $P^+_{l \to r} P^+_{r \to l}$, and $\max[0, m_{r \to l} + m_{l \to r}] = 1$ with probability $P^+_{l \to r}(1 - P^+_{r \to l} - P^-_{r \to l}) + (1 - P^+_{l \to r} - P^-_{l \to r})P^+_{r \to l}$ so that
\begin{align}
  \mathbbm E_\bt \big[ \max [ 0, m_{l \to r} + m_{r \to l} ] \big] &= P^+_{l \to r}(1 - P^-_{r \to l}) + (1 - P^-_{l \to r}) P^+_{r \to l} \,.
\end{align}
We obtain:
\begin{align}
  \mathcal E^*(\bx_1) =& \sum_ l p_l \left[ \prod_{r \in \dd l} P^-_{r \to l} + \prod_{r \in \dd l} \left( 1 - P^+_{r \to l} \right) - 1 \right] + \sum_r \left[ \prod_{l \in \dd r} P^-_{l \to r} + \prod_{l \in \dd r} \left( 1 - P^+_{l \to r} \right) - 1 \right] + \nonumber \\
  & +\sum_{(lr)} \left[ P^+_{l \to r}(1 - P^-_{r \to l}) + (1 - P^-_{l \to r}) P^+_{r \to l} \right] .
\end{align}

\subsubsection{Solving for $\bx_1$}

In order to compute
\begin{align}
  \bx_1^* = \argmin_{\bx_1} \mathbbm E_\bt \min_{\bx_2} \mathcal E(\bx_1, \bt, \bx_2)
\end{align}
we introduce the MS messages in terms of the cavity marginals
\begin{align}
  \Psi_{l_1 \to r} &= \log \frac {\pr_{l_1 \to r} \left[ P^+_{l_1 \to r} = 1 \right]}{\pr_{l_1 \to r} \left[ P^-_{l_1 \to r} = 1 \right]} = \frac {\pr_{l_1 \to r} \left[ x_{l_1 r} = 1 \right]}{\pr_{l_1 \to r} \left[ x_{l_1 r} = 0 \right]} \\
  \Psi_{l_2 \to r}(P^+, P^-) &= \log \pr_{l_2 \to r} \left[ P^+_{l_2 \to r} = P^+, P^-_{l_2 \to r} = P^- \right] + C_{l_2 \to r}
\end{align}
and similarly for $\Psi_{r \to l_1}$ and $\Psi_{r \to l_2}(P^+, P^-)$. The messages $\Psi_{l_1 \to r}$ and $\Psi_{r \to l_1}$ are real numbers, while the messages $\Psi_{l_2 \to r}$ and $\Psi_{r \to l_2}$ are funtions with domain $\{(P^+,P^-) \in [0,1]^2 : P^+ + P^- \leq 1\}$ and codomain $]-\infty, 0]$ (for an appropriate choice of the additive constants $C_{l_2 \to r}$ and $C_{r \to l_2}$). For numerical purposes we shall approximate each of these functions with an array of negative real numbers corresponding to finite size bins for the values of $(P^+, P^-)$ in $[0,1]^2$.

Let us now derive the update equation for $\Psi_{l_1 \to r}$. As in the two-valued case, we need to consider both the vertex and the edge energy contributions. Since $P^+_{l_1 \to r} = P^+_{r \to l_1} = 1 - P^-_{l_1 \to r} = 1 - P^-_{r \to l_1} = x_{l_1 r}$, we only need to consider $P^+_{l_1 \to r}$ for the variables on the outgoing edge and $P^+_{r' \to l_1}$ for the variables on the incoming ones. Also, we remind that $p_{l_1} = 1$ and we obtain:
\begin{align}
  &\mathcal E_{l_1 \to r}(P^+_{l_1 \to r}, \{ P^+_{r' \to l_1}, r' \in \dd l_1 \sm r\}) \nonumber \\
  &\hspace{33mm}= p_{l_1} \left[ \prod_{r \in \dd l_1} P^-_{r \to l_1} + \prod_{r \in \dd l_1} \left( 1 - P^+_{r \to l_1} \right) - 1 \right] + \left[ P^+_{l_1 \to r}(1 - P^-_{r \to l_1}) + (1 - P^-_{l_1 \to r}) P^+_{r \to l_1} \right] \\
  &\hspace{33mm}= 2 (1 - P^+_{l_1 \to r}) \prod_{r' \in \dd l_1 \sm r}(1 - P^+_{r' \to l_1}) - 1 + 2 P^+_{l_1 \to r} \,.
\end{align}
The MS equation is then
\begin{multline}
   \log \pr_{l_1 \to r} \left[ P^+_{l_1 \to r} = P^+ \right] \\
  = \max_{
	\substack {
	  \left\{ P^+_{r' \to l_1} \in \{0,1\},\, r' \in \dd l_1 \sm r \right\} \text{ s.t.}: \\
	  P^+ + \sum_{r' \in \dd l_1 \sm r} P^+_{r' \to l_1} \leq 1
	}
  } \Bigg\{ -\mathcal E_{l_1 \to r}(P^+, \{ P^+_{r' \to l_1}, r' \in \dd l_1 \sm r\}) + \sum_{r' \in \dd l_1 \sm r} \log \pr_{r' \to l_1} \left[ P^+_{r' \to l_1} \right] \Bigg\} \,.
\end{multline}
For $P^+ = 1$ all the incoming $P^+_{r' \to l_1}$ must be 0 and we obtain
\begin{align}
  \log \pr_{l_1 \to r} \left[ P^+_{l_1 \to r} = 1 \right] &= -1 + \sum_{r' \in \dd l_1 \sm r} \log \pr_{r' \to l_1 } \left[ P^+_{r' \to l_1} = 0 \right] ,
\end{align}
while for $P^+ = 0$ the incoming $P^+_{r' \to l_1}$ can either be all 0, or one of them can be 1 and all the other 0:
\begin{multline}
  \log \pr_{l_1 \to r} \left[ P^+_{l_1 \to r} = 0 \right] = \max \Bigg\{ -1 + \sum_{r' \in \dd l_1 \sm r} \log \pr_{r' \to l_1} \left[ P^+_{r' \to l_1} = 0 \right] ,  \\
  1 + \max_{r' \in \dd l_1 \sm r} \Bigg[ \log \pr_{r' \to l_1} \left[ P^+_{r' \to l_1} = 1 \right] - \log \pr_{r' \to l_1} \left[ P^+_{r' \to l_1} = 0 \right] + \sum_{r'' \in \dd l_1 \sm \{r,r'\}} \log \pr_{r' \to l_1} \left[ P^+_{r' \to l_1} = 0 \right] \Bigg] \Bigg\}
\end{multline}
\begin{align}
  \hspace{-24mm}= \log \pr_{l_1 \to r} \left[ P^+_{l_1 \to r} = 1 \right] + \max \left[0, \, \max_{r' \in \dd l_1 \sm r} \Psi_{r' \to l_1} \right]
\end{align}
so that
\begin{align}
  \Psi_{l_1 \to r} = -\max \left[0, \, \max_{r' \in \dd l_1 \sm r} \Psi_{r' \to l_1} \right].
\end{align}

We now turn to the update equation for $\Psi_{l_2 \to r}(P^+, P^-)$. The variables on the outgoing edge, $P^+_{l_2 \to r}$ and $P^-_{l_2 \to r}$, are both in $[0,1]$ and they satisfy $0 \leq P^+_{l_2 \to r} + P^-_{l_2 \to r} \leq 1$. The same is true for the variables in the incoming edges, $\{(P^+_{r' \to l_2}, P^-_{r' \to l_2}),\, r' \in \dd l_2 \sm r\}$, which must also satisfy the constraints (\ref{P+_l_to_r}, \ref{P-_l_to_r})
\begin{align}
  P^+_{l_2 \to r} &= p_{l_2} \prod_{r' \in \dd l_2 \sm r} P^-_{r' \to l_2} \label{constraint_P+}\\
  P^-_{l_2 \to r} &= 1 - p_{l_2} \prod _{r' \in \dd l_2 \sm r} \big( 1 - P^+_{r' \to l_2} \big) \label{constraint_P-}
\end{align}
for given $P^+_{l_2 \to r}$ and $P^-_{l_2 \to r}$.

Again, the variables $P^+_{l_2 \to r}$ and $P^-_{l_2 \to r}$ appear in both a vertex and an edge energy terms, and the energy contribution we need to consider is
\begin{align}
  & \mathcal E_{l_2 \to r} \big( (P^+_{l_2 \to r}, P^-_{l_2 \to r}), \{ (P^+_{r' \to l_2}, P^-_{r' \to l_2}), r' \in \dd l_2 \sm r \} \big) \\
  &\hspace{30mm} = p_{l_2} \left[ \prod_{r' \in \dd l_2} P^-_{r' \to l_2} + \prod_{r' \in \dd l_2} \big( 1 - P^+_{r' \to l_2} \big) - 1 \right] + P^+_{r \to l_2} \big( 1 - P^-_{l_2 \to r} \big) + \big( 1 - P^-_{r \to l_2} \big) P^+_{l_2 \to r} \\
  &\hspace{30mm} = 1 - p_{l_2} + P^+_{l_2 \to r} - P^-_{l_2 \to r}
\end{align}
where we made use of the constraints (\ref{constraint_P+}, \ref{constraint_P-}), and which only depends on the outgoing variables (notice that \emph{a priori} it could also depend on $P^+_{r \to l_2}$ and $P^-_{r \to l_2}$, but does not).

The MS equation is
\begin{align}
  \Psi_{l_2 \to r} (P^+, P^-) &= \max_{
  \substack{
	\{ (P^+_{r' \to l_2}, P^-_{r' \to l_2}) \in [0,1]^2 ,\, r' \in \dd l_2 \sm r \} \text{ s.t.:} \\
	P^+_{r' \to l_2} + P^-_{r' \to l_2} \leq 1 \quad (\forall r' \in \dd l_2 \sm r) \\
	P^+ = p_{l_2} \prod_{r' \in \dd l_2 \sm r} P^-_{r' \to l_2} \\
	P^- = 1 - p_{l_2} \prod_{r' \in \dd l_2 \sm r} \big (1 - P^+_{r' \to l_2} \big)
  }} \left\{ -1 + p_{l_2} - P^+ + P^- + \sum_{r' \in \dd l_2 \sm r} \Psi_{r' \to l_2}(P^+_{r' \to l_2}, P^-_{r' \to l_2}) \right\}
\end{align}
which can be computed efficiently thanks to a method similar to (\ref{efficient_max}). We introduce, with an obvious simplification of notation,
\begin{align}
  G_k(y^+,y^-) &= \max_{\substack{
	\{(z_i^+,z_i^-) \in [0,1]^2,\, i = 1, \dots, k\} \text{ s.t.:} \\
	z_i^+ + z_i^- \leq 1 \quad (\forall i = 1, \dots, k) \\
	y^+ = \prod_{i=1}^k z_i^- \\
	y^- = \prod_{i=1}^k (1 - z_i^+)
  }} \left\{ \sum_{i=1}^k f_i(z_i^+, z_i^-) \right\} \displaybreak[0] \\
  &= \max_{\substack{
	(z_k^+, z_k^-) \in [0,1]^2 \text{ s.t.:} \\
	0 \leq z_k^+ \leq 1 - y^- \\
	y^+ \leq z_k^- \leq 1
  }} \Bigg\{ f_k(z_k^+, z_k^-) + \max_{\substack{
	\{(z_i^+, z_i^-) \in [0,1]^2, \, i=1, \dots, k-1\} \text{ s.t.:} \\
	z_i^+ + z_i^- \leq 1 \quad (\forall i = 1, \dots, k-1) \\
	y^+ / z_k^- = \prod_{i=1}^{k-1} z_i^- \\
	y^- / (1 - z_k^+) = \prod_{i=1}^{k-1} (1 - z_i^+)
  }} \left[ \sum_{i_1}^{k-1} f_i(z_i^+, z_i^-) \right] \Bigg\} \displaybreak[0] \\
  &=  \max_{\substack{
	(z_k^+, z_k^-) \in [0,1]^2 \text{ s.t.:} \\
	0 \leq z_k^+ \leq 1 - y^- \\
	y^+ \leq z_k^- \leq 1
  }} \left\{ f_k(z_k^+, z_k^-) + G_{k-1}\left(\frac{y^+}{z_k^-}, \frac{y^-}{1 - z_k^+}\right) \right\}
\end{align}
which is easily computed iteratively starting with
\begin{align}
  G_0(y^+,y^-) &=
	\begin{dcases}
	0 & \text{ if } y^+ = y^- = 1 \\
	-\infty & \text{ otherwise} .
	\end{dcases}
\end{align}

The update equation for $\Psi_{r \to l_1}$ is obtained in a similar way. We notice that the outgoing edge is connected to a deterministic vertex $l_1$, so we must have $P_{l_1 \to r}^+ = P_{r \to l_1} = 1 - P_{l_1 \to r} = 1 - P_{r \to l_1} \in \{0,1\}$ (and we can express all of them in terms of $P_{l_1 \to r}^+$), while the incoming edges are in part connected to deterministic vertices $l'_1 \in \dd r \sm l_1$ (and the corresponding variables satisfy the same relations as those on the outgoing edge) and in part connected to stochastic vertices $l_2 \in \dd r$ (and the corresponding variables are in $[0,1]$).

The energy terms to be considered are again a vertex and an edge term,
\begin{align}
  & \mathcal E_{r \to l_1}\big(P_{r \to l_1}^+, \, \{P_{r \to l'_1}^+, \, l'_1 \in \dd r \sm l_1\}, \, \{(P_{l_2 \to r}^+, P_{l_2 \to r}^-), \, l_2 \in \dd r\} \big) \\
  &\hspace{39mm} =  \Big[ \prod_{l' \in \dd r} P_{l' \to r}^- + \prod_{l' \in \dd r} (1 - P_{l' \to r}^+) - 1 \Big] + \Big[ P_{r \to l_1}^+ (1 - P_{l_1 \to r}^-) + (1 - P_{r \to l_1}^-) P_{l_1 \to r}^+ \Big] \\
  &\hspace{39mm} = (1 - P_{r \to l_1}^+) \prod_{l' \in \dd r \sm l_1} P_{l' \to r}^- + (1 - P_{r \to l_1}^+) \prod_{l' \in \dd r \sm l_1} (1 - P_{l' \to r}^+) - 1 + 2 P_{r \to l_1}^+ \, .
\end{align}
Notice that, again, this only depends on the ``right'' messages: $P_{r \to l_1}^+$ on the outgoing edge (but not $P_{l_1 \to r}^+$), and $P_{l' \to r}^\pm$ on the incoming ones (but not $P_{r \to l'}^\pm$). The messages on the edges connected to $L_1$ vertices are subject to the matching constraint
\begin{align}
  P_{r \to l_1}^+ + \sum_{l'_1 \in \dd r \sm l_1} P_{l'_1 \to r}^+ \leq 1
\end{align}
while the messages on the edges connected to $L_2$ vertices (i.e. $P_{l_2 \to r}^\pm$) are unconstrained, since the only constraints they are subject to are the BP update equations (\ref{P+_l_to_r}, \ref{P-_l_to_r}) that define them in terms of the messages $P_{r \to l_2}^\pm$, and these do not appear in the expression of the energy.

The MS equation will then be:
\begin{multline}
  \log \pr_{r \to l_1}\left[P_{r \to l_1}^+ = 1\right] = \max_{\substack{
	\{(P_{l_2 \to r}^+, P_{l_2 \to r}^-) \in [0,1]^2, \, l_2 \in \dd r\} \text{ s.t.:} \\
	P_{l_2 \to r}^+ + P_{l_2 \to r}^- \leq 1 \quad (\forall l_2 \in \dd r)
  }} \Bigg\{ -1 + \sum_{l'_1 \in \dd r \sm l_1} \log \pr_{l'_1 \to r}\left[ P_{l'_1 \to r}^+ = 0\right] + \\
  + \sum_{l_2 \in \dd r} \log \pr_{l_2 \to r}\left[ P_{l_2 \to r}^+ , P_{l_2 \to r}^- \right] \Bigg\} \, ,
\end{multline}
\begin{multline}
  \log \pr_{r \to l_1}\left[P_{r \to l_1}^+ = 0\right] = \max \Big[ A\Big( \big\{\pr_{l' \to r}\left[P_{l' \to r}^+,P_{l' \to r}^-\right],\, l' \in \dd r \sm l_1 \big\} \Big),\, \\
  B\Big( \big\{\pr_{l' \to r}\left[P_{l' \to r}^+,P_{l' \to r}^-\right],\, l' \in \dd r \sm l_1 \big\} \Big) \Big]
\end{multline}
with
\begin{multline}
  A\Big( \big\{\pr_{l' \to r}\left[P_{l' \to r}^+,P_{l' \to r}^-\right],\, l' \in \dd r \sm l_1 \big\} \Big) \\
  = \max_{\substack{
	\{(P_{l_2 \to r}^+, P_{l_2 \to r}^-) \in [0,1]^2, \, l_2 \in \dd r\} \text{ s.t.:} \\
	P_{l_2 \to r}^+ + P_{l_2 \to r}^- \leq 1 \quad (\forall l_2 \in \dd r)
  }} \Bigg\{ 1 - \prod_{l_2 \in \dd r} P_{l_2 \to r}^- - \prod_{l_2 \in \dd r} (1 - P_{l_2 \to r}^+) + \\
  + \sum_{l'_1 \in \dd r\sm l_1} \log \pr_{l'_1 \to r}\left[ P_{l'_1 \to r}^+ = 0\right] + \sum_{l_2 \in \dd r} \log \pr_{l_2 \to r}\left[ P_{l_2 \to r}^+ , P_{l_2 \to r}^- \right] \Bigg\} \, ,
\end{multline}
and
\begin{multline}
   B\Big( \big\{\pr_{l' \to r}\left[P_{l' \to r}^+,P_{l' \to r}^-\right],\, l' \in \dd r \sm l_1 \big\} \Big) \\
   = \max_{\substack{
	 \{(P_{l_2 \to r}^+, P_{l_2 \to r}^-) \in [0,1]^2, \, l_2 \in \dd r\} \text{ s.t.:} \\
	 P_{l_2 \to r}^+ + P_{l_2 \to r}^- \leq 1 \quad (\forall l_2 \in \dd r)
  }} \Bigg\{ 1 + \max_{l'_1 \in \dd r \sm l_1} \bigg[ \log \pr_{l'_1 \to r}\left[P_{l'_1 \to r}^+ = 1\right] - \log \pr_{l'_1 \to r}\left[P_{l'_1 \to r}^+ = 0\right] \bigg] + \\
  + \sum_{l'_1 \in \dd r \sm l_1} \log \pr_{l'_1 \to r}\left[ P_{l'_1 \to r}^+ = 0\right] + \sum_{l_2 \in \dd r} \log \pr_{l_2 \to r}\left[ P_{l_2 \to r}^+ , P_{l_2 \to r}^- \right] \Bigg\} \,.
\end{multline}
Notice that all the maximisations are trivial, except the one appearing in $A$, which can nonetheless be computed efficiently exploiting its associativity with a method similar to those explained before: we introduce
\begin{align}
  G_k(y^+, y^-) &= \max_{\substack{
	\{(z_i^+, z_i^-) \in [0,1]^2, \, i = 1, \dots, k\} \text{ s.t.:} \\
	y^+ = \prod_{i=1}^k z_i^- \, , \\
	y^- = \prod_{i=1}^k (1 - z_i^+)
  }} \left\{ - \prod_{i=1}^k z_i^- - \prod_{i=1}^k (1 - z_i^+) + \sum_{i=1}^k f_i(z_i^+, z_i^-) \right\} \displaybreak[0] \\
  &= \max_{\substack{
	(z_k^+, z_k^-) \in [0,1]^2 \text{ s.t.:} \\
	z_k^+ + z_k^- \leq 1
	y^+ \leq z_k^- \leq 1 \\
	y^- \leq 1 - z_k^+ \leq 1
  }} \left\{ -y^+ + \frac{y^+}{z_k^-} - y^- + \frac{y^-}{1 - z_k^+} + f_k(z_k^+, z_k^-) + G_{k-1}\left( \frac{y^+}{z_k^-}, \frac{y^-}{1 - z_k^+} \right) \right\} \label{efficient_max2}
\end{align}
and compute it iteratively starting with
\begin{align}
  G_0(y^+, y^-) =
	\begin{dcases}
	  0 & \text{ if } y^+ = y^- = 1 \\
	  -\infty & \text{ otherwise} \, .
	\end{dcases}
\end{align}

Finally, let us derive the update equation for $\Psi_{r \to l_2}(P_{r \to l_2}^+, P_{r \to l_2}^-)$. Some of the incoming edges will be connected to deterministic vertices $l_1$, with variables in $\{0,1\}$ satisfying the normalization constraint $P_{l_1 \to r}^+ + P_{l_1 \to r}^- = 1$ and the matching constraint $\sum_{l_1 \in \dd r} P_{l_1 \to r}^+ \leq 1$. The remaining incoming edges will be connected to stochastic vertices $l'_2$, with continuous variables satisfying the normalization constraint $P_{l'_2 \to r}^+ + P_{l'_2 \to r}^- \leq 1$ and the constraint
\begin{align}
  P_{r \to l_2}^+ &= \prod_{l \in \dd r \sm l_2} P_{l \to r}^-\, , \\
  P_{r \to l_2}^- &= 1 - \prod_{l \in \dd r \sm l_2} \left( 1 - P_{l \to r}^+ \right)
\end{align}
derived from the update equations (\ref{P+_r_to_l}, \ref{P-_r_to_l}) and involving the variables connected to \emph{all} the incoming edges (both stochastic and deterministic).

The energy term contains both a vertex and an edge contribution and is given by
\begin{align}
  & \mathcal E_{r \to l_2} \big( P_{r \to l_2}^+, P_{r \to l_2}^-, \{P_{l_1 \to r}^+, l_1 \in \dd r, \{(P_{l'_2 \to r}^+, P_{l'_2 \to r}^-), \, l'_2 \in \dd r \sm l_2\} \big) \nonumber \\
  &\hspace{48mm} = \prod_{l \in \dd_r} P_{l \to r}^- + \prod_{l \in \dd r} \left( 1 - P_{l \to r}^+ \right) - 1 + P_{l_2 \to r}^+ \left( 1 - P_{r \to l_2}^- \right) + \left( 1 - P_{l_2 \to r}^- \right) P_{r \to l_2}^+ \\
  &\hspace{48mm} = P_{r \to l_2}^+ - P_{r \to l_2}^-
\end{align}
which only depends on the variables on the outgoing edge.

We can now write the MS equation as
\begin{multline}
  \log \pr_{r \to l_2}[P^+, P^-] = \max_{\substack{
	\big\{ P_{l_1 \to r}^+ \in \{0, 1\}, \, l_1 \in \dd r \big\} \, , \\
	\big\{ \big(P_{l'_2 \to r}^+, P_{l'_2 \to r}^-\big) \in [0,1]^2 , \, l'_2 \in \dd r \sm l_2 \big\} \text{ s.t.:} \\
	\sum_{l_1 \in \dd r} P_{l_1 \to r}^+ \leq 1 , \\
	\prod_{l \in \dd r \sm l_2} P_{l \to r}^- = P^+, \\
	1 - \prod_{l \in \dd r \sm l_2} \big( 1 - P_{l \to r}^+ \big) = P^-
  } } \Bigg \{ P^+ - P^- + \sum_{l_1 \in \dd r} \log \pr_{l_1 \to r}[P_{l_1 \to r}^+] + \\
  + \sum_{l'_2 \in \dd r \sm l_2} \log \pr_{l'_2 \to r}[P_{l'_2 \to r}^+, P_{l'_2 \to r}^-] \Bigg\} \, .
\end{multline}
When $P^+ > 0$, the constraint $P^+ = \prod_{l \in \dd r} P_{l \to r}^-$ forces \emph{all} the $P_{l_1 \to r}^+$ to be 0, and the equation simplifies as
\begin{multline}
  \log \pr_{r \to l_2}[P^+, P^-] = \max_{\substack{
	\big\{ \big(P_{l'_2 \to r}^+, P_{l'_2 \to r}^-\big) \in [0,1]^2 , \, l'_2 \in \dd r \sm l_2 \big\} \text{ s.t.:} \\
	\prod_{l'_2 \in \dd r \sm l_2} P_{l'_2 \to r}^- = P^+, \\
	1 - \prod_{l'_2 \in \dd r \sm l_2} \big( 1 - P_{l'_2 \to r}^+ \big) = P^-
  } } \Bigg \{ P^+ - P^- + \sum_{l_1 \in \dd r} \log \pr_{l_1 \to r}[P_{l_1 \to r}^+ = 0] + \\
  + \sum_{l'_2 \in \dd r \sm l_2} \log \pr_{l'_2 \to r}[P_{l'_2 \to r}^+, P_{l'_2 \to r}^-] \Bigg\}
\end{multline}
which is again of the form (\ref{efficient_max2}).

If instead $P^+ = 0$, the constraint $P^+ = \prod_{l \in \dd r \sm l_2} P_{l \to r}^-$ can be satisfied by setting to 0 either one of the $P_{l_1 \to r}^-$ (and only one, because of the matching constraint), or at least one of the $P_{l'_2 \to r}^-$. In the first case, the corresponding $P_{l_1 \to r}^+$ will be 1, and the constraint $P^- = 1 - \prod_{l \in \dd r \sm l_2} \left( 1 - P_{l \to r}^+ \right)$ can be satisfied only if $P^- = 1$. In the second case on the other hand $P_{l'_2 \to r}^-$ is a continuous variable and $P^-$ can be smaller than 1. We then have
\begin{multline}
  \log \pr_{r \to l_2}[0, P^-] = \max_{\substack{
	\big\{ \big(P_{l'_2 \to r}^+, P_{l'_2 \to r}^-\big) \in [0,1]^2 , \, l'_2 \in \dd r \sm l_2 \big\} \text{ s.t.:} \\
	\prod_{l'_2 \in \dd r \sm l_2} P_{l'_2 \to r}^- = 0, \\
	1 - \prod_{l'_2 \in \dd r \sm l_2} \big( 1 - P_{l'_2 \to r}^+ \big) = P^-
	} } \Bigg \{ - P^- + \sum_{l_1 \in \dd r} \log \pr_{l_1 \to r}[P_{l_1 \to r}^+ = 0] + \\
	+ \sum_{l'_2 \in \dd r \sm l_2} \log \pr_{l'_2 \to r}[P_{l'_2 \to r}^+, P_{l'_2 \to r}^-] \Bigg\} \,.
\end{multline}
(a special case of the previous equation) and
\begin{align}
  \log \pr_{r \to l_2}[0, 1] &= \max \Big[ A \big( \big\{ \pr_{l \to r}[P_{l \to r}^+, P_{l \to r}^-],\, l \in \dd r \sm l_2 \big\} \big),\, B \big( \big\{ \pr_{l \to r}[P_{l \to r}^+, P_{l \to r}^-],\, l \in \dd r \sm l_2 \big\} \big) \Big]
\end{align}
with
\begin{multline}
  A \big( \big\{ \pr_{l \to r}[P_{l \to r}^+, P_{l \to r}^-],\, l \in \dd r \sm l_2 \big\} \big) = \max_{
	\big\{ \big(P_{l'_2 \to r}^+, P_{l'_2 \to r}^-\big) \in [0,1]^2 , \, l'_2 \in \dd r \sm l_2 \big\}
	} \Bigg \{ -1 + \\
	+ \sum_{l_1 \in \dd r} \log \pr_{l_1 \to r}[P_{l_1 \to r}^+ = 0] + \sum_{l'_2 \in \dd r \sm l_2} \log \pr_{l'_2 \to r}[P_{l'_2 \to r}^+, P_{l'_2 \to r}^-] \Bigg\} \,
\end{multline}
and
\begin{multline}
  B \big( \big\{ \pr_{l \to r}[P_{l \to r}^+, P_{l \to r}^-],\, l \in \dd r \sm l_2 \big\} \big) = \max_{
	\big\{ \big(P_{l'_2 \to r}^+, P_{l'_2 \to r}^-\big) \in [0,1]^2 , \, l'_2 \in \dd r \sm l_2 \big\}
	} \Bigg\{ -1 \\
	+ \max_{l_1 \in \dd r} \bigg[ \log \pr_{l_1 \to r}[P_{l_1 \to r}^+ = 1] - \log \pr_{l_1 \to r}[P_{l_1 \to r}^+ = 0] \bigg] + \sum_{l_1 \in \dd r} \log \pr_{l_1 \to r}[P_{l_1 \to r}^+ = 0] + \\
	+ \sum_{l'_2 \to r \in \dd r \sm l_2} \log \pr_{l'_2 \to r}[P_{l'_2 \to r}^+, P_{l'_2 \to r}^-] \Bigg\} \,.
\end{multline}
Notice that the maximisations in $A$ and $B$ are unconstrained, and therefore immediate.

Despite their appearence, these update equations are implemented straightforwardly. To improve the convergence properties of the algorithm, we added both a noise and a reinforcement term to the messages on the deterministic edges (as explained for the two-valued case).

\subsection{Numerical results}

In order to validate our approach, we performed three series of numerical tests. First, we compared the results of the two- and three-valued versions of the algorithm. As we shall see, both algorithms give solutions with energies that are very close to each other both for small and large connectivities, the main difference between the two algorithms beeing the running time. Second, we compared the performance of the two-valued algorithm with a  greedy heuristic based on the average of $p(\bt)$. We shall see that the two-valued algorithm finds solutions with significantly smaller energy when $c > e$. Third, we compared the performance of the two-valued algorithm with the standard method used to solve two-stage optimization problems: stochastic programming. We find that stochastic programming has an acceptable running time for $c < e$ (but still takes $\sim \hspace{-1.5mm} 100$ times longer than the two-valued algorithm to find a solution with the same energy), while for $c > e$ its time performance worsens dramatically, and it becomes practically impossible to solve instances with $|L_1| = 1\,000$ vertices and $c = 3.0$.

In all three cases we did extensive numerical simulations to compare the performance of the different algorithms, both in terms of the energy of the solution obtained and in terms of the running time (and, crucially, its scaling with the size of the system).

\subsubsection{Comparison between the two- and three-valued results}

As a first test, we did a series of comparisons between the results of the two- and three-valued versions of the algorithms on the same set of instances. As we mentioned previously, the three-valued version has a running time which is much longer than the two-valued version, so we did this comparison on relatively small-sized instances, with $|L_1| = 300$ and $|L_2| = |R| = 600$. Since the probabilities $p_{l_2}$ are drawn uniformly in $[0,1]$, the typical ``final'' instance is roughly balanced, with $|L| \simeq 600 = |R|$. We used both reinforcement and noise as described previously, with parameters $\rho = 0.001$ and $\bar \eta = 0.001$ and with a number of bins $B = 10$ to discretize the one- and two-dimensional distributions. The values of these parameters were chosen based on a separate series of comparative runs with several values of $\rho$, $\bar \eta$ and $B$ on the same ensemble of instances. The convergence criterion used in the presence of reinforcement is that the values of the $\bx_1$ variables do not change for $I = 100$ iterations. Each data point is computed as the average over $\simeq 350$ instances. For each instance the energy is computed by extracting a sample of $\mathcal S = 300$ realizations of the stochastic parameters $\bt$, computing the optimal $\bx_2^*$  for each realization, and averaging the corresponding energy over the sample. The total running time with these parameters for typical instances is of the order of 1 second for the two-valued algorithm and a few minutes for the three-valued one, and we obtained convergence for all the instances we tried.
The average energies obtained in these runs are shown in Table \ref{tab:2_vs_3}.

\begin{table}[h]
  \centering
  \begin{tabular}{c c c c}
    \hline \hline
    $c$ \quad & \quad Two-valued \quad & \quad Three-valued \quad & \quad Difference\\
    \hline
    $ 2.0 $ \quad & \quad $ 274.94 \pm 0.54 $ \quad & \quad $ 275.16 \pm 0.54 $ \quad & \quad $ 0.78 \pm 0.69 $ \\
    $ 2.5 $ \quad & \quad $ 190.84 \pm 0.47 $ \quad & \quad $ 190.96 \pm 0.47 $ \quad & \quad $ 0.88 \pm 0.72 $ \\
    $ 3.0 $ \quad & \quad $ 129.42 \pm 0.42 $ \quad & \quad $ 129.50 \pm 0.41 $ \quad & \quad $ 0.72 \pm 0.57 $ \\
    $ 3.5 $ \quad & \quad $  84.95 \pm 0.32 $ \quad & \quad $  85.02 \pm 0.32 $ \quad & \quad $ 0.76 \pm 0.62 $ \\
    $ 4.0 $ \quad & \quad $  54.58 \pm 0.25 $ \quad & \quad $  54.54 \pm 0.25 $ \quad & \quad $ 0.75 \pm 0.60 $ \\
    $ 5.0 $ \quad & \quad $  25.69 \pm 0.19 $ \quad & \quad $  25.87 \pm 0.18 $ \quad & \quad $ 0.89 \pm 1.02 $ \\
    $ 6.0 $ \quad & \quad $  18.86 \pm 0.18 $ \quad & \quad $  19.26 \pm 0.17 $ \quad & \quad $ 1.84 \pm 1.69 $ \\
    \hline \hline
  \end{tabular}
  \caption{Comparison of the average energy obtained with the two- and three-valued algorithms on the same set of instances. The average connectivity of vertices in $L$ is $c$. The last column gives the average value and standard deviation of the single-sample absolute value of the difference between the two energies. We see that this difference is smaller than $1\%$ for $c \leq 3.5$, smaller than $4\%$ for $c = 4.0$ and $c = 5.0$, and becomes relatively large for $c = 6.0$, when the energy itself is very small.}
  \label{tab:2_vs_3}
\end{table}

As expected, the average energy is exactly the same for $c < e$. In fact, in runs without reinforcement (in which the messages converge to finite values), we also verified that for $c < e$ almost all the elements of the three-valued messages that represent states with $P_{L \to R}^0 \neq 0$ or $P_{R \to L}^0 \neq 0$ are equal to minus infinity, which means that the corresponding cavity marginals are concentrated on the states described by the two-valued fixed points. In fact, for $c = 2.0$ the average number of message elements corresponding to $P_{L \to R}^0 \neq 0$ or $P_{R \to L}^0 \neq 0$ and with finite values is $0.45 \pm 0.01$ (out of 45 matrix elements), the average value of these finite fields being $-42.2 \pm 1.0$, while for $c = 2.5$ these averages are respectively $1.85 \pm 0.03$ and $-14.0 \pm 0.3$.

What is more surprising is that the two-valued algorithm gives results that are almost identical to the three-valued one also for $c > e$. Even in this case we have verified that a small number of message elements corresponding to $P_{L \to R}^0 \neq 0$ or $P_{R \to L}^0 \neq 0$ are different from minus infinity: their average number is between $3.5$ and $5.4$ (depending on $c$, and out of 45 matrix elements), and their average value is between $-6.8 \pm 0.2$ and $-12.9 \pm 0.4$. This means that also for $c > e$ the deviation from a two-valued distribution is small, and helps to explain why the two-valued algorithm has such a good performance. However, for $c > e$ some instances do \emph{not} converge without reinforcement, so this conclusion is only limited to those instances for which convergence is obtained even without reinforcement (between 37\% and 99\% of the instances, depending on $c$).

It is not clear to us why the three-valued solution displays these features, and in particular whether this is a sign that RSB does \emph{not} occur in this ensemble of instances. Anyhow, since the energy obtained with the two-valued algorithm is so good, and its running time is much shorter than for the three-valued one, we have used the two-valued algorithm for all the other tests, both for small and large connectivities.

\subsubsection{Comparison with the greedy heuristic}

We consider the following greedy heuristic. Given an instance of the problem, specified by the graph $G=(L, R; E)$ with $L$ partitioned in $L_1$ and $L_2$ and by the probabilities $\mathbf p = \{p_{l_2}, \, l_2 \in L_2\}$, we assign the first stage variables $\bx_1$ by solving the maximum-weight matching problem with graph $G$ and with weights $w_i$ on the vertices $i \in L \cup R$ equal to 1 for the vertices $l_1 \in L_1$ and $r \in R$, and equal to the probabilities $p_{l_2}$ for vertices $l_2 \in L_2$. To keep the notation as similar as possible to the previous one, we can state the problem as follows:
\begin{align}
  \bx_1^\mathrm{greedy} &= \argmin_{\bx_1} \min_{\bx_2} \left\{ \sum_{l_1 \in L_1} \mathbbm 1 \left[ \sum_{r \in \dd l_1} x_{l_1 r} = 0 \right] + \sum_{l_2 \in L_2} p_{l_2} \mathbbm 1 \left[ \sum_{r \in \dd l_2} x_{l_2 r} = 0 \right] + \sum_{r \in R} \mathbbm 1 \left[ \sum_{l \in \dd r} x_{l r} = 0 \right] \right\}
\end{align}
subject to the matching constraints (\ref{matching}).
Once $\bx_1^\mathrm{greedy}$ is found, we compute the average energy as in the previous paragraph by sampling over 300 realizations of the stochastic parameters $\bt$ and finding the optimal $\bx_2^*$ corresponding to each realization, and then averaging the energy.
As a lower bound to the optimal energy, we also consider the offline optimum obtained with full prior knowledge of $\bt$.

\begin{figure}
  \includegraphics{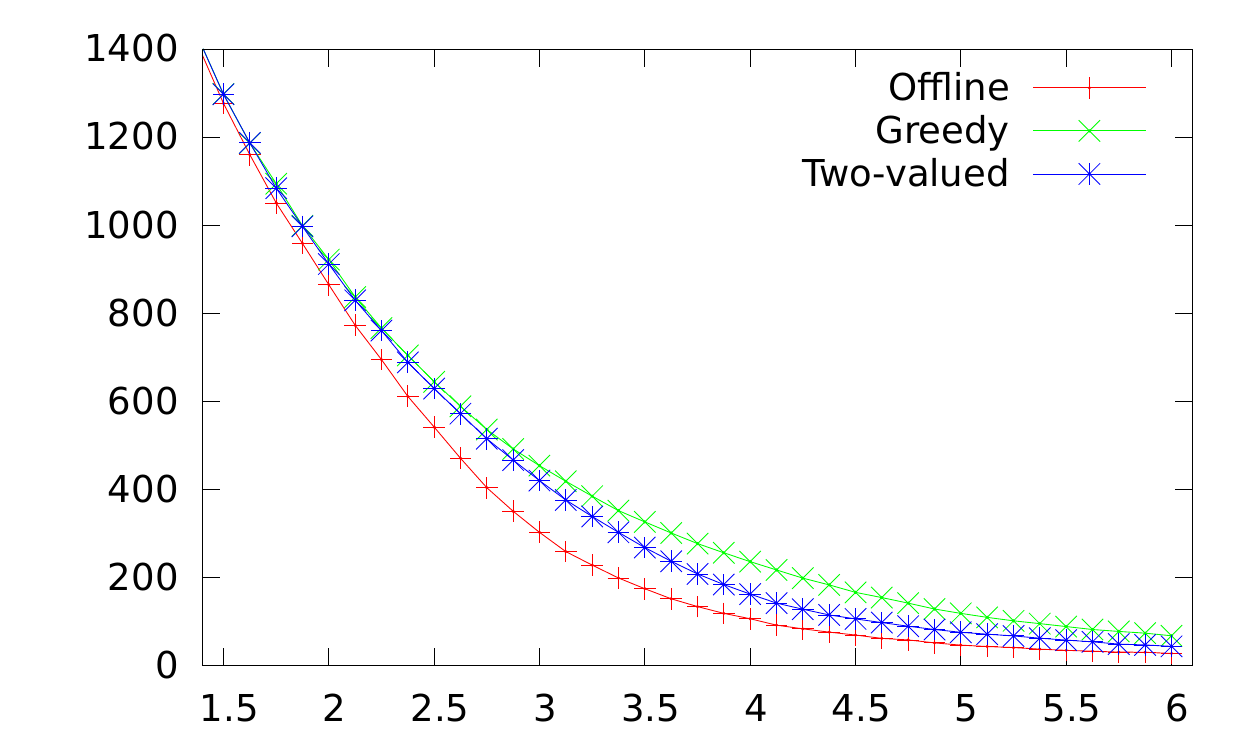}
  \caption{Average energy of the two-valued algorithm vs. the greedy heuristic and the offline optimum. The error bars are smaller than the symbols.}
  \label{fig:2-states_vs_greedy}
\end{figure}

We have compared the results of the two-valued algorithm with the greedy heuristic and the offline optimum for an ensemble of instances with $|L_1| = 1\,000$ and $|L_2| = |R| = 2\,000$. As before, the value of the reinforcement parameter is $\rho = 0.001$ and the value of the noise parameter is $\bar \eta = 0.001$, but the number of bins is increased to $B = 30$ to improve the numerical accuracy (we verified that while there is a small improvement of the energy going from $B = 10$ to $B = 30$, there is almost no further improvement going to $B = 100$). The number of iterations with constant $\bx_1$ required as a convergence criterion is $I = 300$. Each data point is an average of $50$ to $100$ instances (depending on $c$). The total running time of the two-valued algorithm with these parameters on this ensemble of instances is typically less than 1 minute. The results of these simulations are shown in Figure \ref{fig:2-states_vs_greedy}.

We find that the two-valued algorithm always succeeds in finding a solution with a smaller energy than the greedy heuristic. For $c < e$ the difference between the two is very small, and the greedy heuristic is in fact very close to the lower bound for the optimum obtained from the offline solution. As $c$ approaches $e$ the gap between the greedy heuristic and the two-valued algorithm increases, and it becomes larger than $50\%$ of the energy of the two-valued algorithm for $c > 4$.

\subsubsection{Comparison with Stochastic Programming}

Having verified that a simple greedy heuristic fails to provide close-to-optimal solutions for the problem, we have compared the performance of the two-valued algorithm with the standard  technique in the field: stochastic programming. This technique consists in extracting $\mathcal S$ realizations $\{ \bt^1,\dots,\bt^\mathcal S \}$ of the stochastic parameters from the distribution $p(\bt)$ and then observing that
\begin{align}
\min_{\bx_1} \sum_{\bt} p(\bt) \min_{\bx_2} \mathcal E(\bx_1, \bt, \bx_2) &\simeq
\min_{\bx_1} \frac{1}{\mathcal S}\sum_{s=1}^{\mathcal S} \min_{\bx^s_2} \mathcal E(\bx_1, \bt^s, \bx^s_2) \label{eq:stochprog1} \\
 &= \frac{1}{\mathcal S}\min_{\bx_1, \bx^1_2, \dots, \bx^{\mathcal S}_2} \sum
_{s=1}^{\mathcal S} \mathcal E(\bx_1, \bt^s, \bx^s_2) \label{eq:stochprog2}
\end{align}
and the last problem is a standard offline optimization problem that can be
solved using OR techniques like linear relaxations complemented with
branch-and-bound. This approach suffers generally from two separate drawbacks:
one is the approximation in (\ref{eq:stochprog1}) and the second is
that the minimization problem in (\ref{eq:stochprog2}) is NP-Complete \cite{Kong}.

We employed two well known tools for this task: iLog CPLEX, a commercial,
industrial-strenght linear/integer programming software from IBM, and
\texttt{lp\_solve}, an open source alternative. Although qualitatively similar,
results with \texttt{lp\_solve} were uniformly worse than the ones of CPLEX, so
we will not report them.

\begin{figure}
\includegraphics[width=0.6\columnwidth]{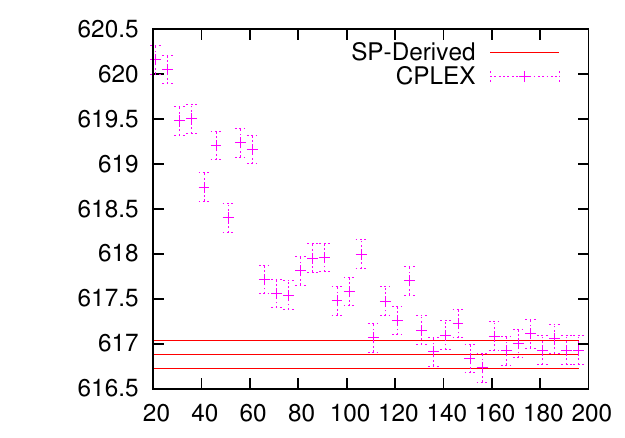}
\caption{Average energy of the solution obtained with stochastic programming as a function of the number of samples $\mathcal S$ for a single instance with $c=2.5$ and $L_1=1\,000$, $|L_2| = |R| = 2\,000$. For each value of $\mathcal S$, the energy of the solution obtained for $\bx_1$ has been computed by resampling over $10\,000$ realizations of $\bt$, finding the optimal $\bx_2$ corresponding to each realization, and averaging the corresponding energies. The horizontal lines correspond to the average energy obtained with the two-valued algorithm and its error bar.}
\label{fig:samples}
\end{figure}

\begin{figure}
\includegraphics[width=0.6\columnwidth]{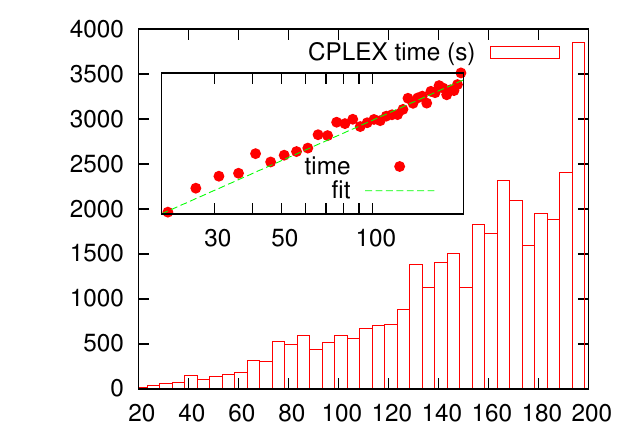}
\caption{CPLEX running  times (in seconds) as a function of the number of samples for the same instance of Figure \ref{fig:samples}. As a comparison, the running time of the two-valued algorithm is of 60 seconds on the same computer for this instance.}
\label{fig:times}
\end{figure}

We observe that the results obtained with stochastic programming depend strongly on $\mathcal S$ and on the average degree $c$. As expected, for fixed $c$ the quality of the solution improves as $\mathcal S$ increases (Figure \ref{fig:samples}), but the running time becomes larger (Figure \ref{fig:times}).

\begin{figure}
\includegraphics[width=0.6\columnwidth]{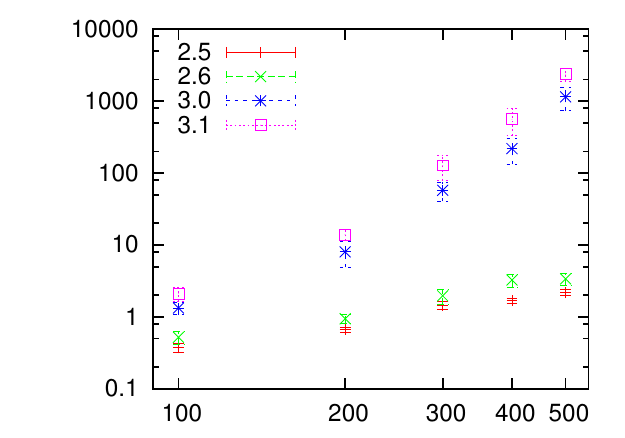}
\caption{CPLEX running times (in seconds) as a function of $|L_1|$ for $\mathcal S = 10$ and four different values of $c$. There seems to be some transition in this behaviour loosely around $c\simeq e$.\label{fig:scaling}}
\end{figure}

For $c < e$, CPLEX seems to be able to solve the problem in polynomial time in both $\mathcal S$ and $N$, but either it is much slower than the SP-derived algorithm or it gives a significantly higher energy (depending on $\mathcal S$). For $c > e$, the time scaling of CPLEX worsens significantly: for $\mathcal S = 10$, the running time increases dramatically with $|L_1|$ (Figure \ref{fig:scaling}), and for $|L_1|=1\,000$ CPLEX was not able to attain an optimum under a cutoff of 24 hours even for $\mathcal S = 2$.

\section{conclusion}\label{S6}
We discussed  the  technical details  which arise  in the generalization of the  message passing algorithm introduced in Ref. \cite{prl}.
In particular, we applied the general scheme to the stochastic maximum weight independent set problem and  to stochastic matching problems.
Extensive numerical comparisons with  local search algorithm based on sampling, linear programming methods and greedy algorithms corroborate 
the idea that the message-passing approach is a valuable alternative to such the traditional techniques.
As a concluding remark we should mention that  the method described in this work is in fact  not limited to stochastic optimization problems.   
There are lots of relevant problems in which one is interested in optimizing a cost function that is 
hard to compute, for example an entropy function or a free energy. Our approach to stochastic optimization problems 
could also be adapted to study these issues.   

\acknowledgments
RZ acknowledges the ERC  grant  OPTINF  267915.  The support from the EC grant STAMINA 265496 is also acknowledged by FA, AB and RZ.

\end{document}